\documentclass[final,3p,times]{elsarticle}
\usepackage{graphicx} 
\usepackage{epsfig}
\usepackage{amssymb}
\usepackage{amsmath}
\usepackage{bm}
\biboptions{square,compress}

\begin{document}

\begin{frontmatter}

\title{Do biological molecular machines act as Maxwell's demons?}

\author{Michal Kurzynski\corref{cor}}
\cortext[cor]{Corresponding author}
\ead{kurzphys@amu.edu.pl}
\author{Przemyslaw Chelminiak\corref{}}

\address{Faculty of Physics, A. Mickiewicz University, Umultowska 85, 61-614 Poznan, Poland}

\begin{abstract}

The nanoscopic isothermal machines are not only energy but also information transducers. We show that the generalized fluctuation theorem with information creation and entropy reduction can be fulfilled for the enzymatic molecular machines with the stochastic dynamics, which offers a choice of the work performance in a variety of ways. A model of such dynamics, specified by a critical complex network, is investigated. The main conclusion of the study is that the processing of free energy has to be distinguished from the processing of organization, which we identify with an adequately defined thermodynamic variable. Maxwell's demon utilizes entropy reduction for creation of information, which, from the former point of view, may be used for a reduction of energy losses, hence ultimately, for the performance of work. From the latter point of view, however, it can be used for other purposes, for example molecular recognition. This can be the case of biological molecular machines. From the biological perspective, the ascertainment is important, that the information creation and storage take place in the long lasting transient stages before completing the free energy transduction cycles. From a broader physical perspective, a supposition could be of special importance, that information is a change of organization, the thermodynamic function of state of the system.

\end{abstract}

\begin{keyword}
Thermodynamics far from equilibrium \sep Fluctuation theorem \sep Native protein dynamics \sep  Critical complex networks


\PACS 05.70.Ln \sep 87.15.H- \sep 87.15.Ya \sep 89.75.Hc


\end{keyword}

\end{frontmatter}


\section{Introduction}

In the intention of its creator \cite{Maxw71}, Maxwell's demon was thought to be an intelligent being, able to perform work at the expense of the entropy reduction of a closed operating system. The perplexing notion of the demon's intelligence was formalized in terms of memory and information processing by Szilard \cite{Szil29}, Landauer \cite{Land61} and subsequent followers \cite{Leef03,Maru09,Saga13}, who pointed out that, in order for the total system to obey the second law of thermodynamics, the entropy reduction should be compensated for by, at least, the same entropy increase, related to the demon's information gain on the operating system's state.

The present, almost universal consensus on this issue is expressed in terms of the feedback control \cite{Saga10,Ponm10,Saga12,Saga13a,Hart14,Horo14}. First, information is transferred from the operating system to memory in the process of measurement (observation) and next, when the system is externally loaded, the transfer of information from memory to the operating system controls the work performance process. The both processes can occur simultaneously \cite{Mand12,Bara14,Horo13,Shir15,Shir15a,Shir16}. Following Landauer's principle, the memory content must be erased at the expense of some entropy production. It should be stressed that the information transfer may \cite{Deff13} but in general is not related to energy transfer between the operating system and memory.

A non-informational formulation of the problem was proposed by Smoluchowski \cite{Smol12} and popularized by Feynman \cite{Feyn63} as the ratchet and pawl machine. It can operate only in agreement with the second law, at the expense of an external energy source \cite{Jarz99}. A. F. Huxley \cite{Huxl57} and consequent followers \cite{Howa01,Howa10} adopted this way of thinking to suggest numerous ratchet mechanisms for the protein molecular machines' action, but no entropy reduction takes place for such models \cite{Jarz99}, thus, they do not act as Maxwell's demons. 

More general models of protein dynamics have been put forward \cite{Kurz98,Fish99,Kolo07,Astu99,Bust01,Lipo00,Lipo09,Kurz03,Kurz14} with a number of intramolecular states organized in a network of stochastic transitions. Here we show that if such models offer work performance in a variety of ways \cite{Kurz14}, the generalized fluctuation theorem \cite{Saga10,Ponm10,Saga12,Saga13a,Hart14,Horo14,Mand12,Bara14,Horo13,
Shir15,Shir15a,Shir16,Jarz11,Seif12,Parr15} is fulfilled with possible entropy reduction. A question appears as to whether they can be considered to act as Maxwell's demons like the artificial nanoscopic machines recenly constructed \cite{Toya10,Beru12,Kosk14,Rold14}. The hypothetical computer model of the network with the Markovian stochastic dynamics is studied, displaying, like networks of the systems biology \cite{Roze10,Esco12}, a transition from the fractal organization on a small length-scale to the small-world organization on the large length-scale.

We start from the general theory of free energy transduction in mesoscopic isothermal machines and the relationship between the entropy and information production. Then, we present the protein molecular machines as isothermal chemo-chemical machines and state a specific model of the protein's stochastic dynamics. A study of this model, using computer simulations, leads us to the result that the free energy transduction in the fluctuating systems must be distinguished from the arrangement transduction. We identify arrangement with an adequately defined thermodynamic variable and show that its increase, related to an entropy reduction and an information creation, takes place in the transient stage before completing the free energy transduction cycle. The latter, in the case of protein machines, can last quite long. Some biological as well as general physical implications of this statement are the subject of the concluding section.

\section{Theory: Formulation of the problem}

\subsection{Stationary isothermal machines}

A long story has been made up for it that the word 'machine' has several different meanings. In our context, we understand a machine to be any physical system that enables two other systems to perform work on one another \cite{Kurz06}. The demon considered by Maxwell changed the temperature to perform work at the expense of absorbed heat. However, the biological molecular machines, like the machine considered by Szilard, operate at a constant temperature. Under isothermal conditions, the internal energy is uniquely divided into the free energy, the component that can be turned into work, and the bound energy (the entropy multiplied by the temperature), the component that can be turned into heat \cite{Call85}. Both the thermodynamic quantities can make sense in the non-equilibrium steady state, if the latter is treated as a partial equilibrium state \cite{Kurz06}. The free energy can be turned irreversibly into the bound energy in the process of internal entropy production, having a meaning of the energy dissipation \cite{Kurz06}. In accordance to such internal energy division, the protein molecular machines are referred to as free energy transducers \cite{Hill89}, thus, when supposed to act as Maxwell's demon, they should perform work at the expense of a part of the dissipation.

\begin{figure}[t]
\centering
\includegraphics[scale=0.75]{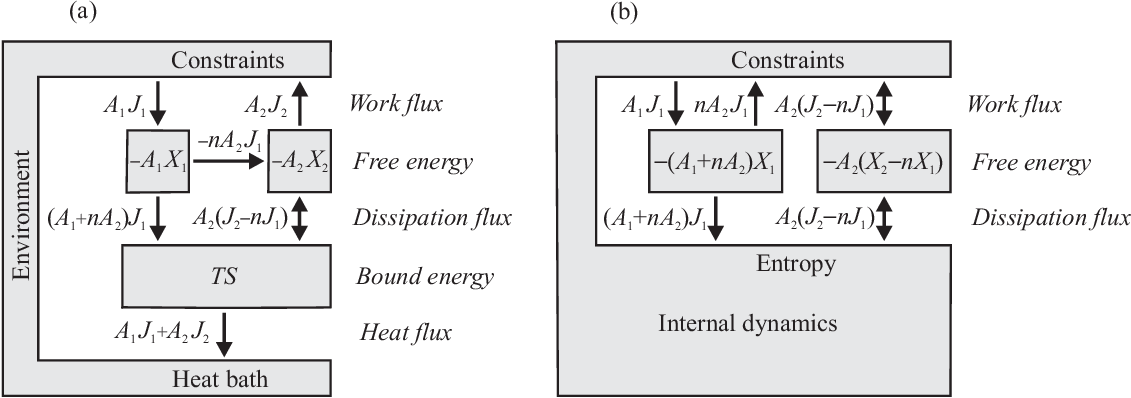}
\caption{Energy processing in the stationary (cyclic) isothermal machine. See the text for the notation explanation. The constraints keep fixed stationary values of the thermodynamic variables $X_i$. We assume these variables to be dimensionless, hence the conjugate forces $A_i$ are of energy dimension and the fluxes $J_i = {\rm d}X_i/{\rm d}t$ are counted in the turnover numbers of the machine per unit time. (a)~The division of the machine's internal energy into the free energy $F = -A_1X_1-A_2X_2$ and the bound energy $TS$. In the steady state, the work flux (the resultant power) equals the dissipation flux, and that equals the heat flux. The transmission ratio $n$ determines the free energy flux from $-A_1X_1$ to $-A_2X_2$. By convention, the fluxes $J_1$ and $J_2$ are assumed to be of the same sign. Then, one system performs work on the other when the forces $A_1$ and $A_2$ are of the opposite sign. The directions of the energy fluxes shown are for $J_1, J_2, A_1 > 0$ and $A_2 < 0$, what is assumed throughout this paper. The actual direction of the flux $A_2(J_2-nJ_1)$, denoted by the forward-reverse arrow, is the subject of this research. (b)~The alternative view of free energy processing in the stationary isothermal machine, determined in the text. Here, both the thermodynamic variables $X_1$ and $X_2-nX_1$ are energetically independent. Only the free energy is specified. The bound energy and the environment are considered to be determined jointly by the internal dynamics of the machine, modified by an interaction with the environment.}
\label{fig01}
\end{figure}

During the stationary isothermal processes, both the free energy and the bound energy remain constant. The energy processing pathways in any stationary isothermal machine are shown in Fig.~\ref{fig01}(a), where the role of all the physical quantities being in use is also indicated. $X_i$ denotes the input ($i=1$) and the output ($i=2$) thermodynamic variable, $A_i$ is the conjugate thermodynamic force and the time derivative, $J_i = {\rm d}X_i/{\rm d}t$, is the corresponding flux. $T$ is the temperature and $S$ is the entropy. The thermodynamic variables $X_1$ and $X_2$ may be mechanical -- displacements, electrical -- charges, or chemical -- numbers of distinguished molecules, hence the fluxes $J_1$ and $J_2$ may be velocities, or electrical or chemical current intensities, respectively. The conjugate forces are then the mechanical forces, or the differences of electrical or chemical potentials (voltages or affinities).

Machines often work as a gear. To clearly specify the transmission ratio $n$ between the fluxes $J_1$ and $J_2$, determining their tight coupling $J_2=nJ_1$, it is important that the thermodynamic variables $X_1$ and $X_2$ be dimensionless. Then, the fluxes are counted in the turnover numbers of the machine per unit time and the forces are of energy dimension. By
convention, the fluxes $J_1$ and $J_2$ are assumed to be of the same sign. Then, one system performs work on the other when the forces $A_1$ and $A_2$ are of the opposite sign. We assume $J_1, J_2, A_1 > 0$ and $A_2 < 0$ throughout this paper.

According to the second law of thermodynamics, the net dissipation flux (the internal entropy production rate, multiplied by the temperature) $A_1J_1+ A_2J_2$ must be nonnegative. However, it consists of two components. The first component, $(A_1+nA_2)J_1$, achieved when the input and output fluxes are tightly coupled, $J_2=nJ_1$, and the output energy flux $A_2J_2=nA_2J_1$ is completely transmitted between the thermodynamic variables $X_1$ and $X_2$ (see Fig.~\ref{fig01}(a)), must also be, according to the same law, nonnegative. Open to discussion is the sign of the complement $A_2(J_2-nJ_1)$ of $(A_1+nA_2)J_1$ to the net dissipation flux $A_1J_1+ A_2J_2$.

In the macroscopic systems, the entropy $S$ is additive and can always be divided into the following two parts $S_1$ and $S_2$, relating to the input and output thermodynamic variables $X_1$ and $X_2$, respectively. As a consequence, the flux $A_2(J_2-nJ_1)$, which corresponds to $S_2$, must also be, under isothermal conditions, nonnegative. This means that, for the assumed negative $A_2$, the output flux $J_2$ should not surpass more than $n$ times the input flux $J_1$. Macroscopically, the second component of the dissipation flux has the obvious interpretation of a slippage in the case of the mechanical machines, a short-circuit in the case of the electrical machines, or a leakage in the case of pumps. However, because of non-vanishing correlations within the bound energy subsystem \cite{Saga10,Ponm10,Saga12,Saga13a,Hart14,Horo14,Mand12,Bara14,Horo13,
Shir15,Shir15a,Shir16,Jarz11,Seif12,Parr15}, in the mesoscopic systems like the protein molecular machines, entropy $S$ is not additive and cannot be divided into two parts like the free energy. This allows the transfer of information within the bound energy subsystem, which could result in the partial reduction of energy dissipation.

In fact, for the biological molecular machines, the output flux $J_2$ can surpass the input flux $J_1$ \cite{Kurz14}. Such a surprising case was observed by Yanagida and his co-workers \cite{Kita97,Kita05}, who found that the single myosin II head can take several steps along the actin filament per ATP molecule hydrolyzed. Whether it changes the sign of the flux $A_2(J_2-nJ_1)$ to the negative depends on establishing the value of the transmission ratio $n$, which, as opposed to the macroscopic machines, is not a simple task in the case of molecular machines, and is one of the main topics of the present paper.

From the point of view of the output force $A_2$, subsystem 1 carries out work on subsystem 2 while subsystem 2 carries out work on the environment. Jointly, the flux of the resultant work (the resultant power) $A_2(J_2-nJ_1)$ is driven by the force $A_2$. The complement to $A_1J_1+A_2J_2$ is the flux $(A_1+nA_2)J_1$ driven by the force $A_1+nA_2$. Consequently, the free energy processing from Fig.~\ref{fig01}(a) can be alternatively presented as in Fig.~\ref{fig01}(b), with the free energy transduction path absent. Here, the subsystems described by two variables $X_1$ and $X_2-nX_1$, respectively, are energetically independent. However, like the subsystems described by the two variables $X_1$ and $X_2$, they are still statistically correlated. Note that in Fig.~\ref{fig01}(b), only the free energy is specified. The bound energy and the environment are considered as determined jointly by the internal dynamics of the machine, modified by an interaction with the environment.

\subsection{Generalized fluctuation theorem}

In the mesoscopic machines, the work, dissipation and heat are fluctuating random variables and their variations, proceeding forward and backward in time, are related to each other by the fluctuation theorem \cite{Jarz11,Seif12,Parr15}. In fact, the feedback control description of the Maxwell's demon action \cite{Saga10,Ponm10,Saga12,Saga13a,Hart14,Horo14}, mentioned in the Introduction, was based on the fluctuation theorem. For the stationary process, the probability distribution function for the input and output fluxes, in general, depending on the time period $t$ of determination, satisfies the stationary fluctuation theorem in the Andrieux-Gaspard form \cite{Andr07,Gasp13}:
\begin{equation}
\label{eq1}
\frac{p(j_{1}(t),j_{2}(t))}{p(-j_{1}(t),-j_{2}(t))}
=\exp\beta[A_{1}j_{1}(t)+A_{2}j_{2}(t)]t \,.
\end{equation}
Here, $\beta = 1/k_{\rm B}T$, where $k_{\rm B}$ is the Boltzmann constant, and $p$ is the joint probability distribution function for the statistical ensemble of the fluxes $j_1(t)$ and $j_2(t)$ over the time period $t$ and their inverses. Eq.~(\ref{eq1}) can be equivalently rewritten as the Jarzynski equality \cite{Jarz97}
\begin{equation}
\label{eq2}
\langle\exp(-\sigma)\rangle=1
\end{equation}
with the stochastic dimensionless entropy production (the energy dissipation divided by  $k_{\rm B}T$)
\begin{equation}
\label{eq3}
\sigma=\sum_{i}\beta A_{i}\mathcal{J}_{i}(t)t \,.
\end{equation}
$\mathcal{J}_{i}(t)$ in (\ref{eq3}) denotes the random variable of the mean net flux over the time period $t$, ($i=1,2$), whereas $j_i(t)$ in (\ref{eq1}) denotes its particular value. $\langle\ldots\rangle$ is the average over the ensemble of the fluxes $j_1(t)$ and $j_2(t)$. Time $t$ must be long enough for the considered ensemble to comprise only stationary fluxes. The convexity of the exponential function provides the second law of thermodynamics:
\begin{equation}
\label{eq4}
\langle\sigma\rangle\geq0
\end{equation}
to be a consequence of (\ref{eq2}). Only the averages of the random fluxes can be identified with the stationary fluxes $J_i$. They are time-independent, $\langle\mathcal{J}_i(t)\rangle = J_i$ for arbitrary $t$. 

In further discussion, for brevity, we will omit the argument $t$ specifying all the fluxes. In the context of the transition from Fig.~\ref{fig01}(a) to (b), the two-dimensional probability distribution function $p(j_1,j_2)$ can be treated as a two-dimensional probability distribution function of two variables $j_1$ and $j_2-nj_1$, with $j_2-nj_1$ as a whole treated as a single variable. If we calculate the marginal probability distributions, then, from the fluctuation theorem (\ref{eq1}) for the total entropy production in both the stationary fluxes $J_1$ and $J_2$, the generalized fluctuation theorems for $J_1$ and the difference $J_2-nJ_1$ follow, respectively, in the logarithmic form:
\begin{equation}
\label{eq5}
\ln\frac{p(j_{1})}{p(-j_{1})} 
= \beta(A_{1}+nA_{2})j_{1}t+\beta A_{2}\langle(\mathcal{J}_{2}
-n\mathcal{J}_{1})t\rangle
- \left<\ln\frac{p(\mathcal{J}_{2} - n\mathcal{J}_{1}\!\mid\!j_{1})}
{p(-\mathcal{J}_{2}+n\mathcal{J}_{1}\!\mid\!-j_{1})}\right>
\end{equation}
(here, the averages $\langle\ldots\rangle$ are taken over the ensemble of the flux differences $j_2-nj_1$) and
\begin{equation}
\label{eq6}
\ln\frac{p(j_{2}-nj_{1})}{p(-j_{2}+nj_{1})} 
= \beta A_{2}(j_{2}-nj_{1})t+\beta(A_{1}+nA_{2}) \langle\mathcal{J}_{1}t\rangle 
- \left<\ln\frac{p(\mathcal{J}_{1}\!\mid\!j_{2}-nj_{1})}
{p(-\mathcal{J}_{1}\!\mid\!-j_{2}+nj_{1})}\right>
\end{equation}
(here, the averages $\langle\ldots\rangle$ are taken over the ensemble of the fluxes $j_1$). Above, we introduced conditional probabilities. The first components on the right of Eqs.~(\ref{eq5}) and (\ref{eq6}) describe the entropy production, now only in the separate fluxes $J_1$ and $J_2-nJ_1$, respectively, but the interpretation of the remaining components is not so easy. The problem is that for the stationary processes, as opposed to the transient nonequilibrium protocols considered in Refs.~\cite{Saga10,Ponm10,Saga12,Saga13a,Hart14}, the notion of the mutual information is not well defined, as it should be exchanged continuously, without any delay \cite{Horo14}.

This disadvantage may, however, be used to determine the value of the transmission ratio $n$. If it is chosen such that the remaining terms in Eq.~(\ref{eq5}) cancel each other out and only the entropy production term remains:
\begin{equation}
\label{eq7}
\ln\frac{p(j_{1})}{p(-j_{1})}=\beta(A_{1}+nA_{2})j_{1}t \,,
\end{equation}
then, Eq. (\ref{eq6}) for the flux $J_2-nJ_1$ can be rewritten in terms of the partly averaged mutual information differences:
\begin{equation}
\label{eq8}
\ln\frac{p(j_{2}-nj_{1})}{p(-j_{2}+nj_{1})}
= \beta A_{2}(j_{2}-nj_{1})t 
- \left<\ln\frac{p(\mathcal{J}_{1},j_{2}-nj_{1})}
{p(\mathcal{J}_{1})p(j_{2}-nj_{1})}\right> 
+ \left<\ln\frac{p(-\mathcal{J}_{1},-j_{2}+nj_{1})}
{p(-\mathcal{J}_{1})p(-j_{2}+nj_{1})}\right> \,.
\end{equation}
Like the antecedent entropic term, both the informative components in Eq. (8) depend only on the single variable $j_2-nj_1$.

The replacement of (\ref{eq5}) by (\ref{eq7}) and (\ref{eq6}) by (\ref{eq8}) corresponds to treating the thermodynamics of the system as bipartite \cite{Horo14}. Although the fully averaged mutual information does not distinguish between the sender from the recipient, the unilaterally averaged informations may differ substantially \cite{Saga13}. In our case, we are dealing with completely asymmetrical coarse graining of the fluxes $\mathcal{J}_1$ and $\mathcal{J}_2-n\mathcal{J}_1$. This means that $\mathcal{J}_1$, when averaged over the ensemble of the flux differences $j_2-nj_1$, is unable to make any choice related to the creation of information, whereas the flux difference $\mathcal{J}_2(t)-n\mathcal{J}_1(t)$, when averaged over the ensemble of the fluxes $j_1$, is able to choose to create some information. In other words, Eq.~(\ref{eq7}) can be considered as the condition for $J_1$ to be a hidden thermodynamic variable, from which and to which the information does not flow \cite{Shir15,Shir15a,Mehl12,Borr15}. The specific value of $n$ results from the internal dynamics of the system; an illustrative example for the biological molecular machine will be presented further on in Fig.~\ref{fig06}, which is a confirmation of the method of the reduction ratio designation that we have chosen.

To conclude, we write the fluctuation theorem for the flux $J_1$, Eq.~(\ref{eq7}), in the form analogous to the Jarzynski equality (\ref{eq2}), from which the second law inequality (\ref{eq4}) follows. However, the Jarzynski equality for the flux $J_2-nJ_1$, Eq.~(\ref{eq8}), should be written in the generalized form \cite{Saga10,Ponm10,Saga12,Saga13a,Hart14,Horo14,
Mand12,Bara14,Horo13,Shir15,Shir15a,Shir16,Jarz11,Seif12,Parr15}:
\begin{equation}
\label{eq9}
\langle\exp(-\sigma-\iota)\rangle=1 \,,
\end{equation}
from which the generalized second law inequality follows:
\begin{equation}
\label{eq10}
\langle\sigma\rangle+\langle\iota\rangle\geq 0 \,.
\end{equation}
As in Eq.~(\ref{eq2}), $\sigma$ has the meaning of the random dimensionless entropy production, while the additional quantity
\begin{equation}
\label{eq11}
\iota = - \ln\frac{p(\mathcal{J}_{1},\mathcal{J}_{2}-n\mathcal{J}_{1})}{p(\mathcal{J}_{1})
p(\mathcal{J}_{2}-n\mathcal{J}_{1})} 
+ \ln\frac{p(-\mathcal{J}_{1},-\mathcal{J}_{2}+n\mathcal{J}_{1})}
{p(-\mathcal{J}_{1})p(-\mathcal{J}_{2}+n\mathcal{J}_{1})} 
\end{equation}
represents the difference of the stochastic information, which the fluctuating flux $\mathcal{J}_2(t)-n\mathcal{J}_1(t)$ sends outside to the flux $\mathcal{J}_1(t)$, when proceeding, respectively, in the forward and backward directions. Accordingly, $\langle\iota\rangle$ has the direct interpretation of the information production by the flux $J_2-nJ_1$. Because the variable $J_1$ is hidden, this information production is, from the point of view of this variable, the information loss, hence positive. This statement remains true for any nanoscopic isothermal machine. Whether the complementing dissipation $\langle\sigma\rangle$ in Inequality (\ref{eq10}) may be negative, it depends on the specific dynamics of the system. Further on, we study this problem for the biological molecular machines.

\subsection{Proteins as chemo-chemical machines}

\begin{figure}[ht]
\centering
\includegraphics[scale=0.6]{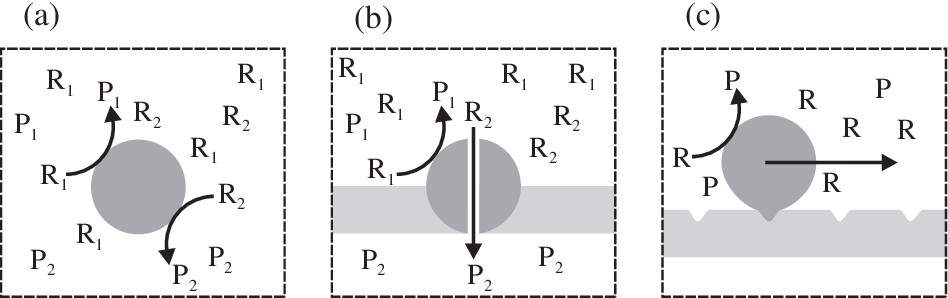}
\caption{A simplified representation of the three types of the biological molecular machines: (a)~enzymes that simultaneously catalyze two reactions, (b)~pumps placed in a biological membrane, and (c)~motors moving along a track. Constraints are symbolized by the frame of the broken line. It shoud be stressed that the entire system inside the frame is the machine: the number of substrate molecules determines its free energy (the thermodynamic state) whereas the enzyme's internal dynamics determines its bound energy (the entropy).}
\label{fig02}
\end{figure}

From a theoretical point of view, it is convenient to treat all biological molecular machines as chemo-chemical machines \cite{Kurz06}. The protein chemo-chemical machines are enzymes, that simultaneously catalyze two effectively unimolecular reactions: the free energy-donating input reaction ${\rm R}_1 \leftrightarrow {\rm P}_1$ and the free energy-accepting output reaction ${\rm R}_2 \leftrightarrow {\rm P}_2$ (Fig.~\ref{fig02}(a)). Also, pumps and molecular motors can be treated in the same manner. Indeed, the molecules present on either side of a biological membrane can be considered to occupy different chemical states (Fig.~\ref{fig02}(b)), whereas the external load influences the free energy involved in binding the motor to its track (Fig.~\ref{fig02}(c)), which can be expressed as a change in the effective concentration of this track \cite{Fish99,Kolo07,Kurz03,Kurz06}.

The system considered consists of a single enzyme macromolecule, surrounded by a solution of its substrates, possibly involving the track (Fig.~\ref{fig02}). It is an open system with constraints controlling the mean numbers of incoming and outgoing molecules and, in particular, the number of steps performed by the motor along the track. Under specified relations between the concentration of the substrates \cite{Kurz14}, two independent stationary (nonequilibrium) molar concentrations of the products [P$_1$] and [P$_2$], related to the enzyme total concentration [E], are to be treated as the input and output dimensionless thermodynamic variables $X_1$ and $X_2$, respectively, presented in Fig.~\ref{fig01}. The fluxes $J_i$ with the conjugate thermodynamic forces $A_i$ are determined as \cite{Kurz06,Hill89}
\begin{equation}
\label{eq12}
J_{i}\equiv\frac{\mathrm{d}}{\mathrm{d}t}X_{i}
=\frac{\mathrm{d}}{\mathrm{d}t}\frac{[\mathrm{P}_{i}]}
{[\mathrm{E}]},\;\;\;\;\;\;\;\;\beta A_{i}=\ln\frac{[\mathrm{P}_{i}]^{\mathrm{eq}}}{[\mathrm{R}_{i}]
^{\mathrm{eq}}}\frac{[\mathrm{R}_{i}]}{[\mathrm{P}_{i}]} ,
\end{equation}
where the superscript eq denotes the corresponding equilibrium concentrations. The fluxes can be calculated if a model of the system's dynamics is at disposal \cite{Kurz03,Kurz14}.

It is now well established that most if not all enzymatic proteins display slow stochastic dynamics of transitions between a variety of conformational substates composing their native state \cite{Kurz98,Henz07,Uver10,Chou11,Chod14,Shuk15}. Two types of experiments imply this statement. The first includes observations of the non-exponential initial stages of a reaction after the special preparation of an initial microscopic state in a statistical ensemble of biomolecules by, e.g., the laser pulse \cite{Aust75,Frau10}. The second type of experiments is imaging and the manipulation of single biomolecules in various processes \cite{Ishi01,Yana08}. The even more convincing proof of the conformational transition dynamics of native proteins has been provided by molecular dynamics simulations \cite{Kita98,Karp02}. Here, as a symbol of progress made recently, the study of conformational transitions in intrinsically disordered native kinases could be quoted, which a few years ago resulted in a network of 25 substates \cite{Yang09} and now offers a network of several hundred items \cite{Hu16}.

\begin{figure}[t]
\centering
\includegraphics[scale=0.5]{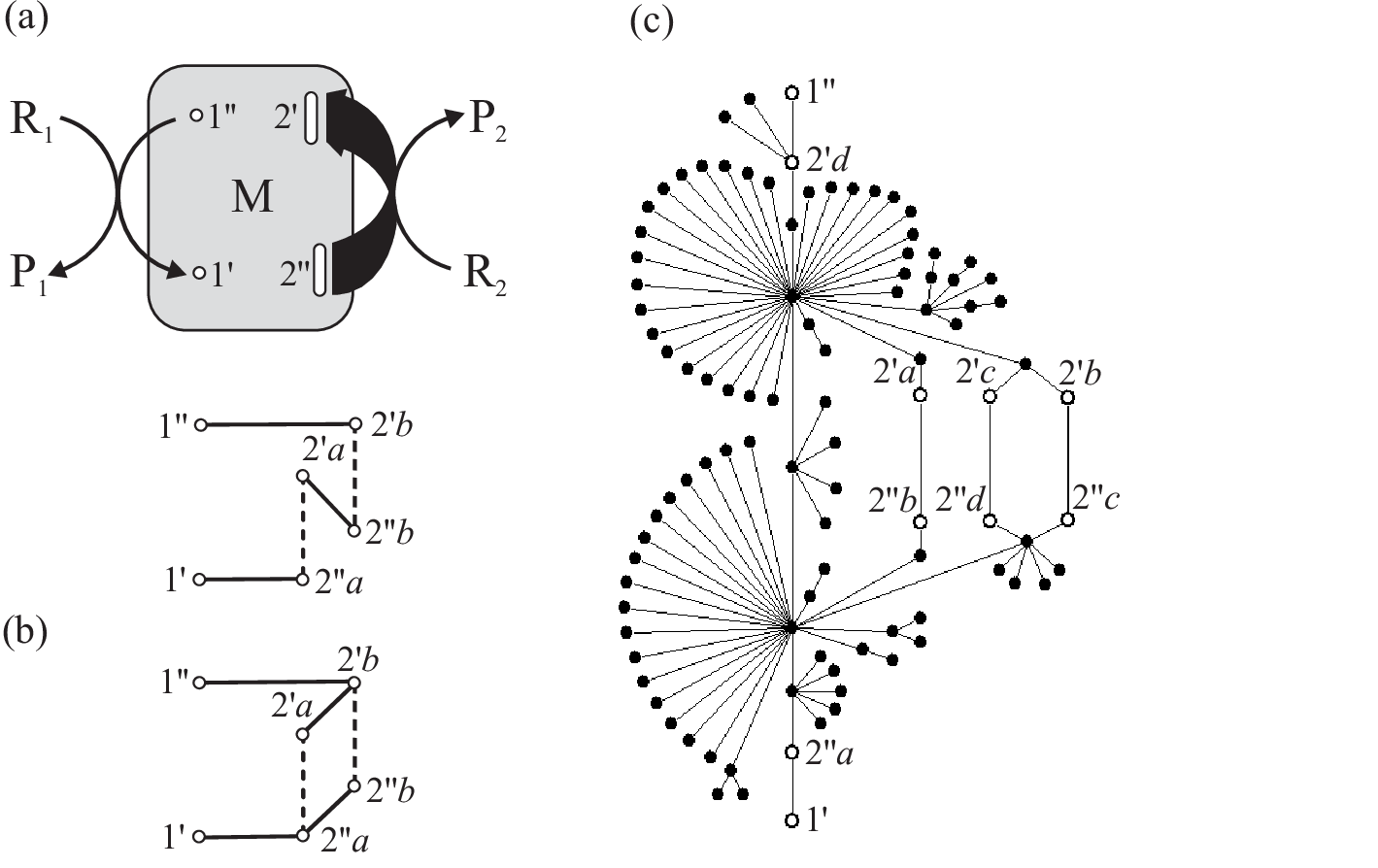}
\caption{The kinetics of the enzymatic chemo-chemical machine. (a)~General scheme. The grey box represents an arbitrary one-component network of transitions between conformational substates, composing either the enzyme or the enzyme-substrates native state \cite{Kurz14}. All these transitions satisfy the detailed balance condition. A single pair (the 'gate') of conformational states $1''$ and $1'$ is distinguished, between which the input reaction ${\rm R}_1 \leftrightarrow {\rm P}_1$ brakes the detailed balance. Also, a single or a variety (ovals) of pairs of conformational states $2''$ and $2'$ is distinguished, between which the output reaction ${\rm R}_2 \leftrightarrow {\rm P}_2$ does the same. All the reactions are reversible; the arrows indicate the directions assumed to be forward. (b)~The simplest examples of the network of transitions with two output gates connected in series (up) or in parallel (down). In the absence of internal transitions indicated by the dashed lines, both the transmission ratio $n=2$ and the degree of coupling $\epsilon=2$ in the serial case, whereas both $n=1$ and $\epsilon=1$ in the parallel case. In the presence of internal transitions, the value of $n$ is lower in the serial case and higher in the parallel case, whereas the value of $\epsilon$ is lower in the both cases. (c)~Exemplifying implementation of the 100-node network, constructed following the algorithm described in Methods. The stochastic dynamics on this network is also described there. Note the two hubs, the states of the lowest free energy, that can be identified with the two main conformations of the protein machine, e.g., 'open' and 'closed', or 'bent' and 'straight', usually the only ones occupied sufficiently high to be notable under equilibrium conditions. The single pair of the output transition states (the gate) chosen for the simulations is $(2''a, 2'd)$. The alternative four output pairs $(2''a, 2'a)$, $(2''b, 2'b)$, $(2''c, 2'c)$ and $(2''d, 2'd)$ are chosen tendentiously to lie one after another.}
\label{fig03}
\end{figure}

It follows that on the mesoscopic level, the dynamics of a specific biological chemo-chemical machine is the Markov process described by a system of master equations, determining a network of conformational transitions that satisfy the detailed balance condition (we mentioned more representative examples in the Introduction \cite{Kurz98,Fish99,Kolo07,Astu99,Bust01,Lipo00,Lipo09,Kurz03,Kurz14}),
and a system of pairs of distinguished nodes (the 'gates') between which the input and output chemical reactions force transitions, that break the detailed balance \cite{Kurz03,Kurz14} (Fig.~\ref{fig03}(a)). Recently, we proved analytically \cite{Kurz14} that, for a single output gate, when the enzyme has no opportunity for any choice, the ratio $\epsilon = J_2/J_1$ cannot exceed one. This case also includes the various ratchet models \cite{Howa01,Howa10}, which assume the output and input gates to coincide. The output flux $J_2$ can
only exceed the input flux $J_1$ in the case of many output gates, that seems to be the rule rather than the exception (the output 'fluctuating reaction rate')  \cite{Kurz98,Zwan90,Lu98,Engl06,Kurz08,Xie13}.

At this point, it is worth explaining demonstratively the relationship between the degree of coupling $\epsilon$ and the transmission ratio $n$, which is essential for determining the direction of the flux $A_2(J_2-nJ_1)$ in Figs.~\ref{fig01}(a) and (b). For the assumed negative force $A_2$, the sign of $A_2(J_2-nJ_1) = A_2(\epsilon - n)J_1$ can be negative only if $\epsilon > n$. The value of the degree of coupling $\epsilon$ depends, in general, on the values of the forces $A_i$ \cite{Kurz03,Kurz14,Kurz06}, whereas the specific value of $n$ results from the topology of the network. Two examples of extremely simple networks with two output gates are presented in Fig.~\ref{fig03}(b). In the absence of internal transitions indicated by dashed lines, for the gates connected in series, both $n=2$ and $\epsilon=2$, independently on the forces $A_i$, whereas for the gates connected in parallel, both $n=1$ and $\epsilon=1$. Consequently, the flux $A_2(J_2-nJ_1) = A_2(\epsilon - n)J_1$ vanishes in both cases, which is obvious, since there is no short circuit of the input gate that omits the output gates. The only difference between both examples is that in the case of the gates connected in series, the system passes along the output gates successively, whereas in the case of the gates connected in parallel, it has a choice which output gate to pass along first. However, by assumption, the network of internal conformational transitions should be one-component, not broken, so that transitions shown by dashed lines must be taken into account. It allows the choice of the output gate in both examples and results in decreasing the value of $\epsilon$ \cite{Kurz14}. Adding yet other possible transitions changes the value of the transmission ratio $n$ to be established in the whole range from 1 to 2.

Determination of the values of $\epsilon$ and $n$ in actual, more complex networks is not a simple task. Our goal in the following is to consider the biological molecular machines with a dynamics with many output gates allowing a choice. Since very poor experimental support is still available for actual conformational transition networks in native proteins, we restrict our attention only to a model network.

\section{Methods: Specification of the computer model}

Various models of the networks with several output gates have been considered \cite{Kurz14}, but one class of models seems to be the most realistic, based on a hypothesis that the protein conformational transition networks, like the higher level biological networks, the protein interaction network and the metabolic network, have evolved in the process of a self-organized criticality \cite{Bak96,Snepp05}. Most networks of the systems biology are scale-free and display a transition from the fractal organization on the small length-scale to the small-world organization on the large length-scale \cite{Roze10,Esco12}.

In Ref.~\cite{Kurz14} we have shown that the case of many different output gates was reached in a natural way on scale-free fractal trees, extended by long-range shortcuts. A network of 100 nodes with such properties is depicted in Fig.~\ref{fig03}(c). The algorithm of constructing the stochastic scale-free fractal trees was adopted after Goh et al. \cite{Goh06}. Shortcuts, though more widely distributed, were considered by Rozenfeld, Song and Makse \cite{Roze10}. Here, we randomly chose three shortcuts from the set of all the pairs of nodes distanced by six links. The network of 100 nodes in Fig.~\ref{fig03}(c) is too small to determine its scaling properties, but a similar procedure of construction applied to $10^5$ nodes results in a scale-free network, which is fractal on a small length-scale and a small world on a large length-scale. Very recently, an actual network, which seems to possess similar properties and comprising some 250 nodes, was obtained in a long, 17 $\mu$s molecular dynamics simulation \cite{Hu16}.

To provide the network with stochastic dynamics, we assumed the probability of changing a node to any of its neighbors to be the same in each random walking step \cite{Kurz14,Chel17}. Then, following the detailed balance condition, the free energy of a given node is proportional to the number of its links (the node degree). The most stable nodes are the hubs, which are the only practically observed conformational substates under equilibrium conditions. For a given node $l$, the transition probability to one of the neighboring nodes is inversely proportional to the number of links, thus to the equilibrium occupation probability $p_l^{\rm eq}$ of the output node, and equals $(p_l^{\rm eq}\tau_{\rm int})^{-1}$, where $\tau_{\rm int}$ is the mean time of internal transition counted in the random walking steps. This time is determined by the doubled number of links minus one \cite{Chel17}, $\tau_{\rm int}=2\dot(100+3-1)=204$ random walking steps for the 100 node tree network with 3 shortcuts assumed. To compare, we found the mean first passage time between the most distant nodes to equal 710 random walking steps.

The forward external transition probability, related to the stationary concentration [P$_i$], is determined by the mean time of external transition $\tau_{\rm ext}$, and equals $(p_i^{\rm eq}\tau_{\rm ext})^{-1}$ per random walking step, $p_i^{\rm eq}$ denoting the equilibrium occupation probability of the initial input or output node ($i=1,2$, respectively). The corresponding backward external transition probabilitiy is modified by the detailed balance breaking factors $\exp (-\beta A_i)$. The assumed mean time of forward external transition $\tau_{\rm ext}=20$ was one order of the magnitude shorter than the mean time of internal transition $\tau_{\rm int}=204$, so that the whole process is controlled by the internal dynamics of the system, not the external. The assumption that most biochemical reactions are controlled by the slow dynamics of the proteins has a very rich literature confirmations \cite{Kurz98,Kurz06}. It is this, in fact, that led to the now widely accepted change of the fundamental paradigm of molecular biology, that not only structure but also dynamics determine the function of the proteins \cite{Henz07,Uver10,Chou11,Chod14,Shuk15,Kurz14a}.

\section{Results}

\subsection{Determination of the mean cycle duration}

All the averages in the equations from (\ref{eq2}) to (\ref{eq10}) are to be performed over a statistical ensemble of the stationary fluxes determined for the time period $t$. We can obtain such an ensemble by dividing a long stochastic trajectory of random walk on the studied network into the segments of the length $t$, and, next, by determining the net numbers of external transitions, hence, the values of the fluxes $j_1(t)$ and $j_2(t)$ for each segment. However, we have to be sure that the time period $t$ is long enough for the considered ensemble to comprise only stationary, but not transient fluxes.

Because the initial state is random in the successive divisions of the trajectory into the segments of equal length, the averaged value of the flux
$\langle\mathcal{J}_i(t)\rangle$ (but not the higher moments!) coincides practically with its stationary value $J_i$, even for the very short time period $t$. To evaluate the actual time, after which the fluxes become stationary, we have first to divide the whole trajectory into the cycles or 'protocols' \cite{Jarz11,Seif12,Parr15} of transient trajectories, starting and ending in the same state, say $1'$. For the ensemble of the cycles of the length $t$, or $t$ within a certain small range, we can determine the time-dependent averages $J_1(t)$ and $J_2(t)$, and in such a way find the mean cycle duration $\tau_{\rm cycle}$, long after which the dynamics of the studied system passes from the transient to the stationary stage.

\begin{figure}[ht]
\centering
\includegraphics[scale=0.35]{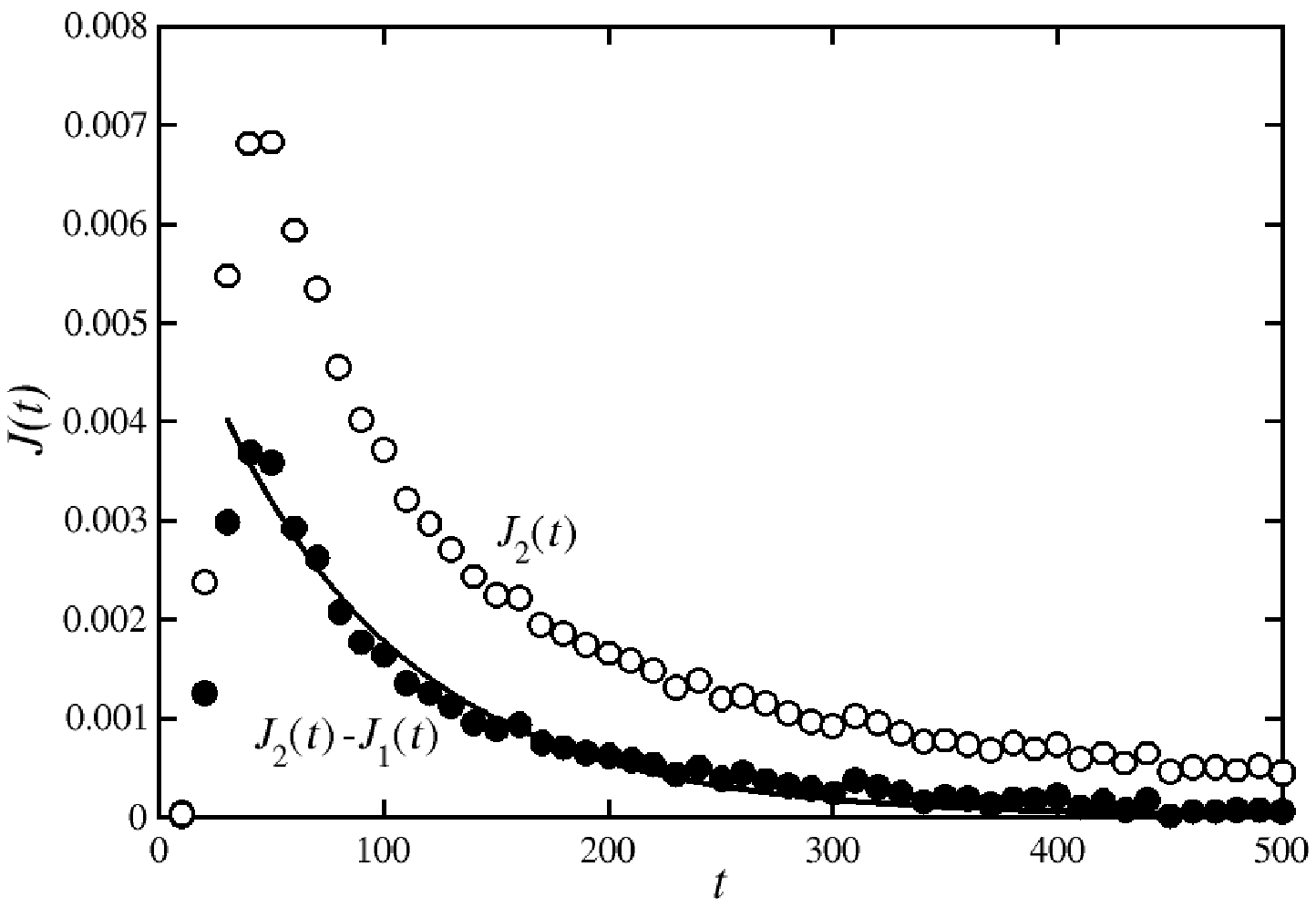}
\caption{The time dependence of the averages over cycles $J_2(t)$ (the white circles) and $J_2(t)-J_1(t)$ (the black circles) found in the model random walk simulations on the network shown in Fig.~\ref{fig03}(c) with the four output gates and the stationary flux ratio $J_2/J_1=1.59$. The solid line corresponds to exponential fitting, with $\tau_{\rm cycle}=100$ random walking steps, of the flux difference $J_2(t)-J_1(t)$ downfall to zero.}
\label{fig04}
\end{figure}

We performed Monte Carlo simulations of random walk in $10^{10}$ computer steps on the network shown in Fig.~\ref{fig03}(c) with the dynamics specified above, with both the single and the fourfold output gate. The gates were chosen tendentiously to maximize the value of the degree of coupling $\epsilon = J_2/J_1$. We assumed $\beta A_1 = 1$ and a few smaller, negative values of $\beta A_2$ determining the ratio $ J_2/J_1$. The presence of external transitions, breaking the detailed balance, makes some computational complications, which are discussed and explained in details elsewhere \cite{Chel17}.

In Fig.~\ref{fig04}, the time dependence of $J_2(t)$ and the difference $J_2(t)-J_1(t)$ for the fourfold output gate and a chosen value of the ratio $J_2/J_1$ is depicted. We do not quote the time dependence of the separate means over cycles $J_1(t)$ and $J_2(t)$, because they both do not tend to zero in the cycle duration possible to designate. This means that they are not determined completely in the transient stages of dynamics (they have a longer memory). Only the difference $J_2(t)-J_1(t)$ tends to zero, which is understandable, because the transition through the input gate each time erases the memory of this difference. Exponentially fitting the downfall of the flux difference $J_2(t)-J_1(t)$ to zero allows us to evaluate the mean cycle duration in the case of the fourfold output gate to be approximately $\tau_{\rm cycle}=100$ random walking steps.

\subsection{Determination of the transmission ratio}

Many trials with divisions of the whole trajectory into segments of different lengths result in the conclusion that

\hspace{\parindent}\parbox[t]{436pt}
{(i)~The obtained two-dimensional probability distribution functions 
$p(j_1(t),j_2(t))$ actually satisfy the Andrieux-Gaspard fluctuation theorem (\ref{eq1}) for $t$ longer than the mean cycle duration $\tau_{\rm cycle}$.}

\begin{figure}[ht]
\centering
\includegraphics[scale=0.45]{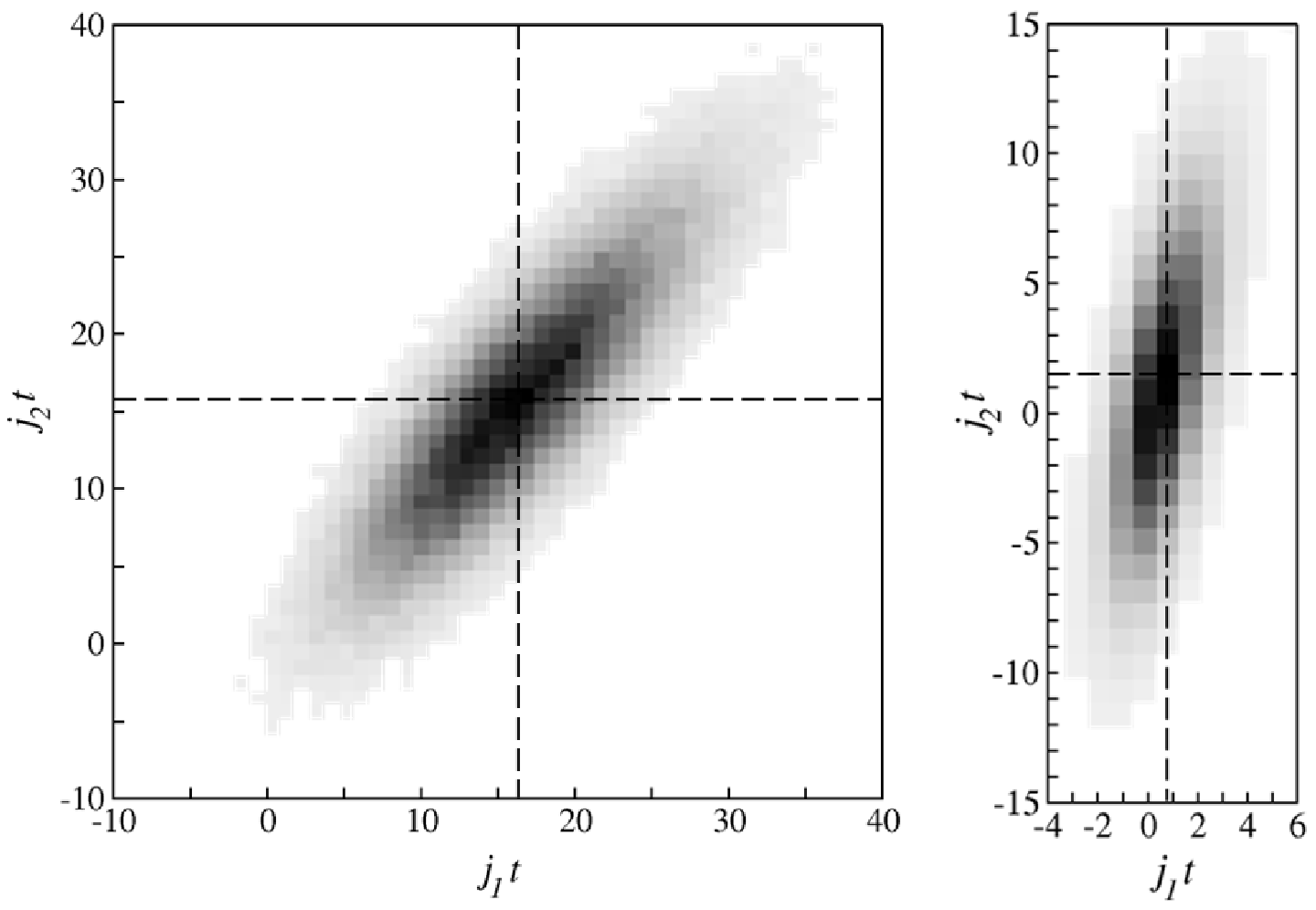}
\caption{Exemplifying two-dimensional probability distribution functions $p(j_1(t),j_2(t))$ found in the model random walk simulations on the network shown in Fig.~\ref{fig03}(c) with the single output gate (left) and the fourfold output gate (right). For the single output gate, we assumed $t = 10^4$ random walking steps and $J_2/J_1 = 0.95$. For the fourfold output gate, we assumed $t = 10^3$ random walking steps and $J_2/J_1 = 1.59$. The averaged values of the individual fluxes, multiplied by the time $t$ of the determination, are marked by the dashed lines.}
\label{fig05}
\end{figure}

\noindent To generate the statistical samples of the stationary fluxes numerously enough, we have chosen the time of the stationary averaging $t=10^4$ random walking steps for the single gate and $t=10^3$ random walking steps for the fourfold gate. Exemplary two-dimensional probability distribution functions $p(j_1,j_2)$ are shown in Fig.~\ref{fig05}.

Next, we calculated the marginals $p(j_1)$ of the two-dimensional probability distribution functions $p(j_1,j_2)$. Upon analyzing the results, we found that

\hspace{\parindent}\parbox[t]{436pt}
{(ii) The logarithm of the ratio of the marginals $p(j_1)/p(-j_1)$ can be described by the formula (\ref{eq5}) with both the entropic and additional components linearly depending on $j_1$. The correction to entropy production is negative.}

\begin{figure}[t]
\centering
\includegraphics[scale=0.5]{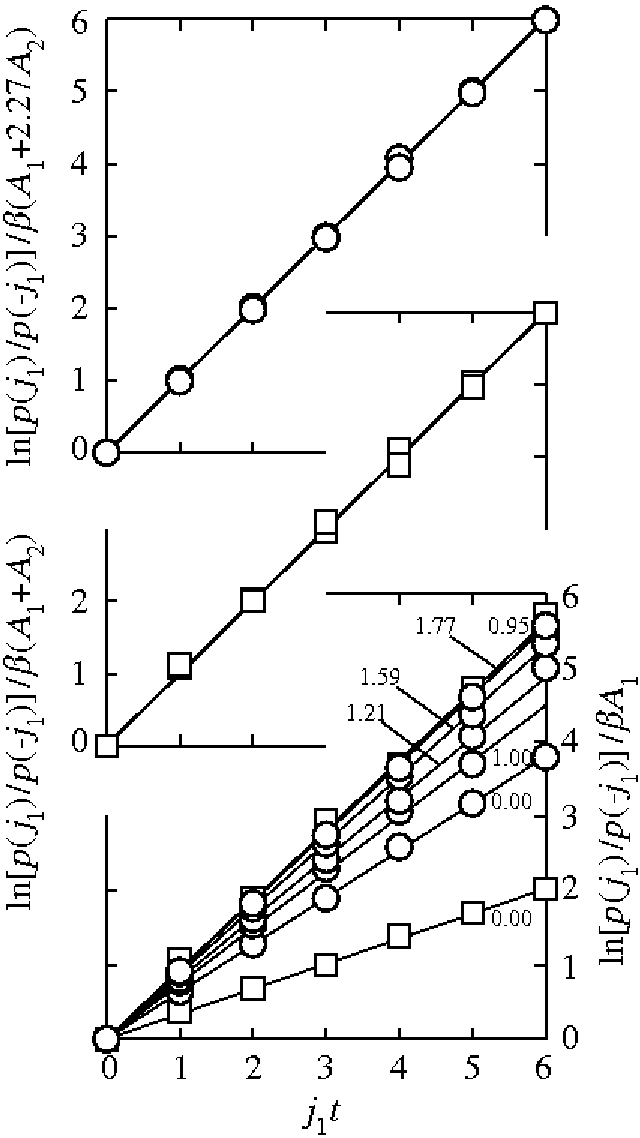}
\caption{The generalized fluctuation theorem dependence (\ref{eq5}) found in the model random walk simulations. The examined network was the one shown in Fig.~\ref{fig03}(c) with the single output gate (the squares) and the fourfold output gate (the circles). We assumed $\beta A_1 = 1$ and a few smaller, negative values of $\beta A_2$ determining the ratio $\epsilon = J_2/J_1$ of the average output and input fluxes noted in the figure. At the beginning, the dependence was related to the force $\beta A_1$ (the lowest diagram). Relating it to $\beta A_1 + nA_2$, Eq.~(\ref{eq7}), allowed us to fit the value of the reduction ratio $n$ to be 1.00 for the single output gate (the higher diagram), and 2.27 for the fourfold output gate (the highest diagram). In the latter two  diagrams, many simulation points practically cover each other, which means that the value of $n$ depends only on the network topology and confirms our method of its designation.} 
\label{fig06}
\end{figure}

\noindent The dependences of $\ln p(j_1)/p(-j_1)$  on $j_1t$, when divided by $\beta A_1$, are plotted in Fig.~\ref{fig06} and fitting these dependences to Eq. (\ref{eq7}) allowed us to determine the value of the transmission ratio to be $n=1.00$ for the single output gate and $n=2.27$ for the fourfold output gate. Let us note, that the determined values of $n$ do not depend on the values of $\beta A_2$, hence are the property of only the assumed topology of the network. 

Knowing the values of $n$, we determined the non-diagonal marginal distributions $p(j_2-nj_1)$ of the two-dimensional probability distributions $p(j_1,j_2)$, and found that

\hspace{\parindent}\parbox[t]{436pt}
{(iii) The logarithm of the ratio of the marginals $p(j_2-nj_1)/p(-j_2+nj_1)$ can be described by the formula (\ref{eq8}) with both the entropic and informative components linearly depending on $j_2-nj_1$. The informative correction is, as expected, positive.}

\begin{figure}[t]
\centering
\includegraphics[scale=0.5]{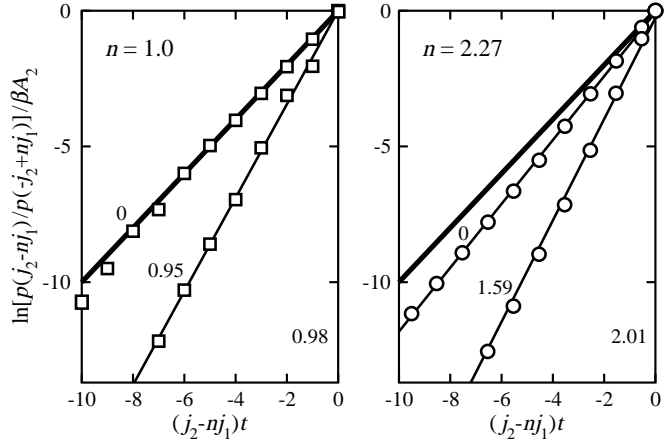}
\caption{The generalized fluctuation theorem dependence (\ref{eq8}) for the marginal $j_2-nj_1$, found in the model random walk simulation for the single ($n=1$, the squares, left) and the fourfold ($n=2.27$, the circles, right) output gate. The thick, solid line corresponds to the lack of information production in Eq. (\ref{eq8}). Three markedly different values of the ratio $\epsilon = J_2/J_1$ were chosen for each case (the vertical line, $\beta A_2 = 0$, corresponds to the maximum value of $\epsilon$ both for the single and the fourfold output gate). Because the corresponding values of $\beta A_2$ are negative, the simulation points lie in the lower half-plane of the graph. See the text for more detailed discussion.}
\label{fig07}
\end{figure}

\noindent Exemplary verifications of Eq. (\ref{eq8}) are depicted in Fig.~\ref{fig07} both for the single (left) and the fourfold (right) output gate. Three markedly different values of the ratio $\epsilon = J_2/J_1$ were chosen for each case. The thick, solid line corresponds to the lack of information production in Eq.~(\ref{eq8}). Such a case takes place for the force $\beta A_2=-0.670$, that stalls the flux through the single output gate ($J_2/J_1=0$). The same effect of stalling the flux through the fourfold output gate is reached for $\beta A_2=-0.173$, but this is accompanied by a non-zero information production (the line with a larger slope) resulting from the possibility of the choice of the output gate even if the resultant net flux is zero. The greatest value of the degree of coupling $\epsilon = J_2/J_1$ is reached for $\beta  A_2=0$, which is represented by the vertical line in both diagrams. It corresponds to $\epsilon=0.98$ for the single output gate, and $\epsilon=2.01$ for the fourfold output gate. As a consequence, because of $\epsilon < n$ in both cases, the entropy production rate 
$\beta A_2(J_2-nJ_1) = \beta A_2(\epsilon-n)J_1$ is always positive. The important conclusion is that the direction of the flux considered in Figs.~\ref{fig01}(a) and (b) to be forward or reverse, should always be forward.

\subsection{Contribution from transient fluxes}

The nanoscopic machines are not only energy but also information transducers. In other words, 'information can serve as a thermodynamic resource similar to free energy' there \cite{Parr15}. Up to then, we treated information as an appropriately averaged mutual information exchanged between two subsystems \cite{Saga10,Ponm10,Saga12,Saga13a,Hart14,Horo14}, but such creation of information at the expense of the entropy reduction appeared not possible in the biological molecular machines. Information may, however, be also considered as the content of a sequence of bits \cite{Mand12,Bara14,Horo13,Shir15,Shir15a}, and the latter  approach is equivalent to the bipartite measurement-control approach \cite{Shir16}. Two types of information in the second sense are produced during the time course of the simulated process with many output gates. The first is in the form of a string with a more or less random sequence of letters, e.g.
\[
\ldots~~b~~c~~d~~d~~a~~c~~b~~b~~a~~b~~c~~c~\ldots \,,
\]
labelling successive gates, which the system passed through. And the second is in the form of a string with a more or less random sequence of signs, e.g.
\[
\ldots~+~+~-~+~-~-~+~+~+~+~\dots \,,
\]
describing the directions (forward or backward) of the successive transitions through the output gates.

Nowadays, it is impossible to determine the information of the first type experimentally, and for its theoretical determination, detailed studies of individual, successive transient protocols related to individual trajectories are needed \cite{Jarz11,Seif12,Parr15}. However, the information of the first type is strongly related to the information of the second type, directly registered in the single biomolecule experiments \cite{Kita97,Kita05,Ishi01,Yana08}. The information of the second type flows out of the machine and defines the number of the product molecules P$_2$, hence, the fluctuating variable $X_2$. Both types of information are erased each time the system passes forward through the input gate $1''1'$ and the molecule P$_1$ is created, hence, the variable $X_1$ increases by one. In other words, the information created by the molecular machine in the successive free energy transduction cycles is written in the fluctuating number of product molecules P$_2$, created per product molecule P$_1$, i.e., in the fluctuating variable $X_2-X_1$ which differs from the variable $X_2-nX_1$ in the case of the multiple output gate. We had already found that the corresponding stationary flux $J_2-J_1$ is determined entirely by the transient stage dynamics (Fig.~\ref{fig04}).

In accordance with the dual interpretation of information \cite{Shir16}, also for the flux $J_2-J_1$, we consider Eq.~(\ref{eq6}) to result in the generalized second law of thermodynamics (\ref{eq10}). Formally, upon substituting $n = 1$, the average dimensionless entropy production (the dissipation $D$ divided by $k_{\rm B}T$) is determined as
\begin{equation}
\label{eq13}
\beta D\equiv\langle\sigma\rangle=\beta A_{2}(J_{2}-J_{1})t
\end{equation}
whereas the average information production $I$, as
\begin{equation}
\label{eq14}
I\equiv\langle\iota\rangle = \beta(A_{1}+A_{2})J_{1}t 
- \left<\ln\frac{p(\mathcal{J}_{1}(t)
\!\mid\!\mathcal{J}_{2}(t)-\mathcal{J}_{1}(t))}
{p(-\mathcal{J}_{1}(t)\!\mid\!-\mathcal{J}_{2}(t)+
\mathcal{J}_{1}(t))}\right> \,.
\end{equation}
The averages are taken over the joint probability function of the fluxes $p(j_1(t),j_2(t))$.

For the macroscopic machines, there is no correlation between the fluxes, so that only the first term in (\ref{eq14}) contributes to the information production $I$. It indirectly represents the entropy production in the hidden variable $X_1$ (in the case of $n=1$, compare Eq.~(\ref{eq7})). For the macroscopic machines as well as the mesoscopic enzymatic machines with the single output gate, the information production $I$ and the entropy production $\beta D$ must always be positive, of the same sign (see Fig.~\ref{fig07}, left). Accordingly, $I$ is to be interpreted as the information loss. It is the general case of the free energy transduction for any value of $n$ (see Fig.~\ref{fig07}, right). 

However, the case of the direct difference $J_2-J_1$ for values of $n$ greater than one is different. In order to determine the temporal fluctuations of the flux $J_2-J_1$ for the fourfold output gate, we calculated the diagonal marginal $p(j_1-j_2)$ of the two-dimensional distributions $p(j_1,j_2)$, and found that

\hspace{\parindent}\parbox[t]{436pt}
{(iv) The logarithm of the ratio of the diagonal marginals $p(j_1-j_2)/p(-j_2+j_1)$ can be described by the formula (\ref{eq6}) with $n = 1$ and all the components linearly depending on the difference $j_2-j_1$. The informative correction is of the opposite sign to dissipation and large.}

\begin{figure}[t]
\centering
\includegraphics[scale=0.5]{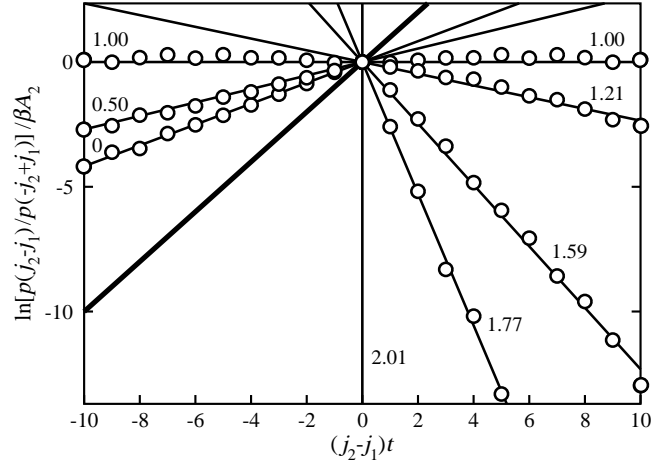}
\caption{The generalized fluctuation theorem dependence (6) with $n=1$, found in the model random walk simulations for the fourfold output gate. The examined network was that shown in Fig.~\ref{fig03}(c). We assumed $\beta A_1 = 1$ and a few negative values of $\beta A_2$ determining the difference $J_2-J_1$ of the stationary output and input fluxes (or the degree of coupling $\epsilon = J_2/J_1$ noted in the figure). The results of each simulation were obtained symmetrically on the whole axis of the fluctuating difference $(j_2-j_1)t$ but in the figure, they are presented, divided by the (negative!) coefficient $\beta A_2$, only on the left from zero for the negative average $J_2-J_1$, and on the right from zero for the positive average $J_2-J_1$. The thick, solid line corresponds to the first, entropic component to the right-hand side of Eq.~(\ref{eq6}). For the multiple output gate, the information gain, resulting from the possibility of a choice, surpasses the information loss in favor of the flux $J_1$, and reduces the effects of the entropy production in part, until the limit of $J_2-J_1 = 0$  ($J_2/J_1 = 1$), above which information production prevails.}
\label{fig08}
\end{figure}

\noindent
The dependences of $\ln p(j_1-j_2)/p(-j_2+j_1)$ on $(j_2-j_1)t$ are depicted in Fig.~\ref{fig08} for a few chosen negative values of $\beta A_2$ that determine the difference of the averages of the output and input fluxes $J_2-J_1$, thus, the ratio of those averages $\epsilon = J_2/J_1$. We discarded the rare results for the values of $(j_2-j_1)t$ higher than 10, as being burdened with a statistical error too large. The simulation data are presented in such a way that the transition from the entropy production to the entropy reduction might be clearly seen. The thick, solid line corresponds to the first, entropic term to the right-hand side of Eq. (\ref{eq6}). For many output gates, as opposed to the case of a single output gate, the information gain, resulting from the possibility of a choice, surpasses the information loss and reduces the effects of the entropy production in part, until the limit of $J_2-J_1 = 0$  ($J_2/J_1 = 1$), above which information production prevails.

\begin{figure}[t]
\centering
\includegraphics[scale=0.35]{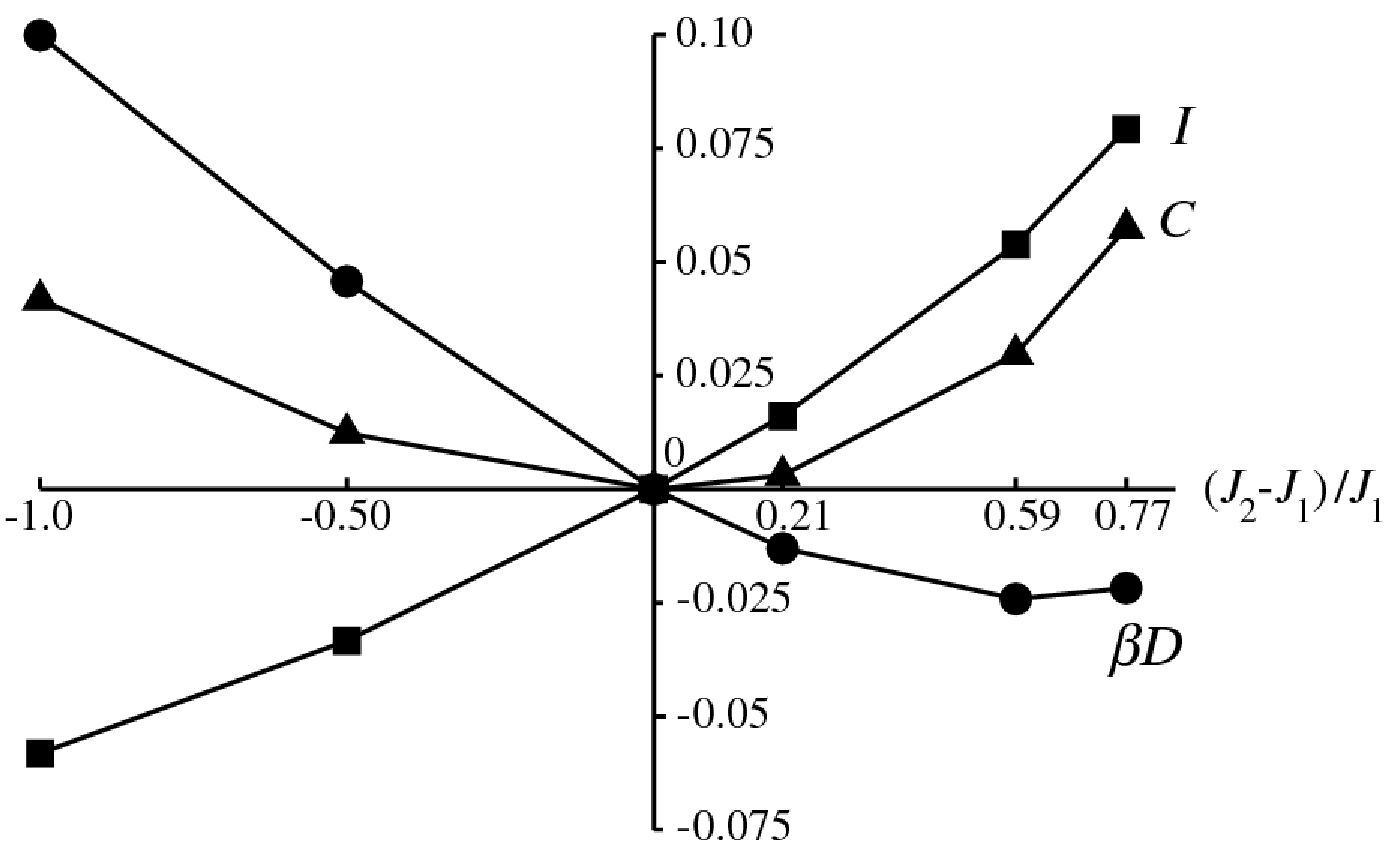}
\caption{The dependence of the information production $I$ (the squares) and the entropy production $\beta D$ (the circles) per $t=10^3$ random walking steps of determination, on the flux difference $J_2-J_1$. The points represent the values obtained from the data given in Fig.~\ref{fig08}. The sum of $\beta D$ and $I$, we refer to as the cost $C$ of information processing (the triangles), is also shown.}
\label{fig09}
\end{figure}

The stationary values of $J_1$ and $J_2$ are unambiguously related to the forces $A_1$ and $A_2$, so we can directly determine the average entropy production $\beta D$, Eq.~(\ref{eq13}), for the given time period $t$. Fig.~\ref{fig08} clearly shows the linear relation
\begin{equation}
\label{eq15}
\beta D+I=\alpha\beta D
\end{equation}
where $\alpha$ is the tangent of the adequate straight line slope. From (\ref{eq15}), knowing $\beta D$ and $\alpha$, we can determine the average information production $I$ without referring to the much more complex formula (\ref{eq14}). The values of $I$ and $\beta D$, obtained from the values of $\alpha$ found from Fig.~\ref{fig08}, are presented in Fig.~\ref{fig09} as the functions of  $J_2-J_1$. Note that for  $J_2-J_1 > 0$, the negative entropy production $\beta D$ is associated with the positive information production $I$, which should now be referred to as actual information creation.

\section{Discussion: Procesing of free energy versus processing of organization}

The fundamental task of statistical thermodynamics is to link the internal dynamics of 'microstates' of a studied system with the dynamics of 'macrostates' -- the thermodynamic variables. In the mesoscopic systems, the thermodynamic variables are fluctuating quantities. In any case, the transition from the 'microstates' to 'macrostates' consists in averaging over time. The change of the availble knowlege of the 'microstate' is refered to as entropy and the change of the available knowledge of the 'macrostate' as information.  

In the present paper, we realized such program for the internal dynamics determined by the complex network presented in Fig.~\ref{fig03} and the macroscopic dynamics in the form of a network connecting two chains of natural numbers, which represent the fluctuating concentrations of reagents $X_1$ and $X_2$ related to a single molecule of protein enzyme. We considered the open system at steady state, so the concentrations $X_1$ and $X_2$ had to be replaced by the corresponding fluxes $J_1$ and $J_2$. Study of the fluctuating flux difference $J_2-J_1$ turned out the most important. It is the time derivative of the thermodynamic variable $X_2-X_1$ that characterizes organization of the system. In the case of the macroscopic mechanical machines, it could be, for example, the possibly slipping angle of a component wheel to the axle. In the case of the macroscopic battery (the electrochemical machine), it is the difference between the available electric charge and the unused amount of reductor, both quantities being determined in molecular units. The organization $X = X_2-X_1$ is controlled by force $A_2$ and can be expressed in the energy units as $A_2X$. The organization of the macroscopic machines always decreases: the wheels slide with respect to the axles, the batteries wear out. This need not be the case of the fluctuating mesoscopic machines.

\begin{figure}[t]
\centering
\includegraphics[scale=0.75]{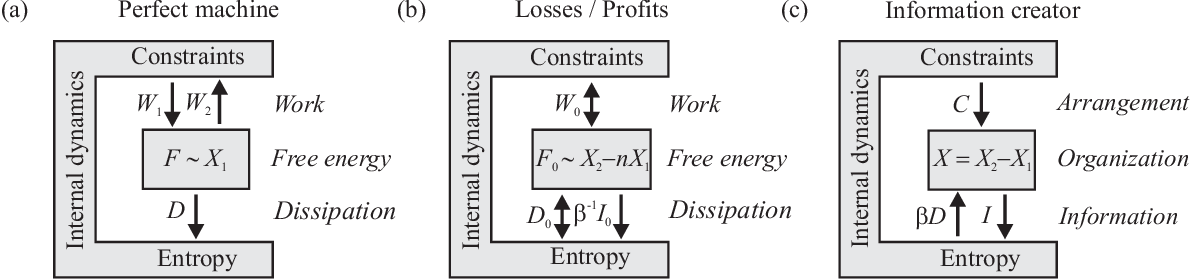}
\caption{Free energy processing versus arrangement processing in the machine participating in a stationary isothermal process. (a)~In the case of the perfect machine, the output flux is tightly coupled to the input flux, hence the free energy of the system is uniquely determined by its input thermodynamic variable $X_1$. (b)~In the actual macroscopic as well as mesoscopic machine, the remaining part of the free energy, proportional to the difference $X_2-nX_1$, 
corresponds to energy losses both for the transmission ratio $n=1$ and greater (see the discussion just after Eq.~(\ref{eq14})). (c)~In the mesoscopic enzymatic machine with many gates, the transmission ratio $n$ is greater than unity and it is the additional variable $X_2-X_1$ that has the meaning of the system's organization. See the text for more detailed discussion. The wishful directions of the input and output work fluxes as well as the input and output information fluxes are indicated. The actual directions of the fluxes denoted by the forward-reverse arrows depend on the specific internal dynamics of the system. In the biological molecular machines with the internal dynamics studied here, these are forward.}
\label{fig10}
\end{figure}

The main result of our study is that the free energy processing has to be distinguished from the organization processing, which is schematically illustrated in Fig.~\ref{fig10}, where we distinguish between the action of the perfect machine and the losses. Note that both free energy of the perfect machine $F$, energy losses $F_0$ and organization $X$ are functions of state of the system. As in Fig.~\ref{fig01}(b), the bound energy and the environment in Fig.~\ref{fig10} are considered to be determined jointly by the internal dynamics of the machine modified by an interaction with the environment. In the case of the purely stochastic dynamics, the presence of the thermal bath influences the probabilities of the internal transitions, whereas the constraints influence the probabilities of the external transitions. 

For the systems operating under isothermal conditions, the first and second laws of thermodynamics:
\begin{equation}
\label{eq16}
W_{1}+W_{2}-\Delta F=D\geq 0 
\end{equation}
and
\begin{equation}
\label{eq17}
W_{0}-\Delta F_{0}=D_{0} + \beta^{-1}I_{0} \geq 0
\end{equation}
are fulfilled for the perfect free energy processing and losses, respectively. For the stationary processes, the free energies $F$ and $F_0$ remain constant, $\Delta F = \Delta F_0 = 0$. All the components of work $W_1$, $W_2$, and $W_0$, dissipation $D$ and $D_0$, and information $I_0$ are functions of the process. The individual components of Eqs. (\ref{eq16}) and (\ref{eq17}), which have been considered in our study, are specified in Fig.~\ref{fig01} and in the text. The free energy $F$ is proportional to $X_1$, and the free energy losses $F_0$ to $X_2-nX_1$. In the perfect machine, the output flux is tightly coupled to the input flux, $J_2=nJ_1$, and the output work is maximum. Dissipation $D$ tends to zero when the perfect stationary machine works infinitely slowly, $J_1 \to 0$, but, then, its input and output powers also drop to zero. Loss in the free energy processing means that no work on the environment is performed despite the forced energy dissipation. We have proven that loss of information $I_0$ occurs in any case of the free energy transduction. In the biological molecular machines also the dissipation $D_0$ is positive. However, this must not be the case of other molecular machines and then, the loss $W_0$ changes into the profit. 

In the mesoscopic enzymatic machines with many gates, hence the possibility of a choise, the loss is determined by the thermodynamic variable $X_2 - nX_1$ with the transmission ratio $n$ greater than unity. However, this variable jointly with $X_1$ does not characterize fully the system. The variable $X_2-X_1$ remains, which now has the meaning of the system's organization. For the processing of organization, the generalized first and second law can be written as
\begin{equation}
\label{eq18}
\beta D + I -\Delta X=C\geq 0 \,.
\end{equation}
Under stationary conditions, the mean organization $X=X_2-X_1$ is constant, $\Delta X = 0$. Here, the two components of the generalized second law (\ref{eq10}), given by Eqs.~(\ref{eq13}) and (\ref{eq14}), are more generally treated as the two components of information. Both $\beta D$ and $I$ as well as $C$, the quantity which balances two former quantities and which could be referred to as the organization cost or arrangement of the system, are functions of the process.

All three changes of organization in Eq.~(\ref{eq18}) are in fact the three types of information and are associated with the three strings of bits. When presenting results of our simulations, we described two signals. The first, in the form of a string of letters, corresponds to the arrangement of organization, whereas the second, in the form of a string of signs, corresponds to the actual information. We did not mention the third signal from the random number generator, which simulates a noise corresponding to entropy production or reduction.

Only the free energy transducer, for which the works $W_1$ and $W_2$ are of the opposite sign, can be referred to as the machine. Similarly, only the organization transducer, for which the information production $I$ and the entropy production $\beta D$ are of the opposite sign, can be referred to as the information creator (see Fig.~\ref{fig10}(c)). For $I$ being positive, $\beta D$ must be negative (the entropy reduction instead of production). The biological molecular machines, for which this is the case, may be said to act as Maxwell's demons, although they do not utilize the information creation for the reduction of energy losses, but only for other purposes, for example molecular recognition.

When considering the coupling of the free energy donating process with the free energy accepting one, the case of $J_2=J_1$ is exceptional. Our simulation clearly points out that for $n > 1$, the case of $J_2-J_1 = 0$ occurs for the total compensation of entropy reduction $\beta D$ by information creation $I$, albeit, on average, both then tend to zero (compare Fig.~\ref{fig09}). This case is the optimum in the sense that the arrangement $C$ of the total process is then zero. We may refer such a situation to as the system's selforganization.

\section{Concluding remarks}

The two powerful theoretical physics tools created at the turn of the century, the fluctuation theorem \cite{Jarz11,Seif12,Parr15} and the idea of self-organized criticality \cite{Bak96,Snepp05}, force a significant change in our views on the nature of the biological molecular machines action. Progress in the theory coincides with the intensive experimental and numerical studies of protein dynamics \cite{Henz07,Uver10,Chou11,Chod14,Shuk15,Aust75,Frau10,
Ishi01,Yana08,Kita98,Karp02,Yang09,Hu16}. 

Under the physiological conditions, the protein molecular machines fluctuate constantly between lots of conformational substates composing their native state. The probabilities of visiting individual substates are far from equilibrium and determined by the concentration of the substrate molecules surrounding the machine. The machine can be considered as an enzyme that simultaneously catalyzes two reactions: the free energy-donating and free energy-accepting ones. During the transient stages before completing the free energy transduction cycles, the subsequent realizations of the free energy-accepting reaction force transitions between separate regions of the conformational substates of the protein machine, which brakes ergodicity \cite{Palm82} of its dynamics until the next free energy-donating reaction begins. The machine transduces information about the types of successive transitions on the information about their successive directions. In this transduction the third signal can be utilized. It is the purely random noise of the environment, which decides the realization of actual transitions between the conformational substates. Information about the direction of the successive transitions during the transient stage of the free energy transduction cycle determines the fluctating difference of the concentrations of the free-energy-accepting and free-energy-donating molecules, surrounding the machine. This difference is a thermodynamic measure of the dynamical organization of the molecular machine  and can be the source of information passed to a further use.

A possible proposal, not to be underestimated, is that the biological structural memory, traditionally associated with DNA and RNA, can in this way be complemented by a dynamical memory of proteins. The transient dynamics stages during the succeeding free energy transduction cycles can last quite long \cite{Hu16,Metz16}, especially when more shortcuts and external transitions in the conformational transition network are taken into account \cite{Kurz14a}. Projection of the trajectory on the fluctuating thermodynamic variable of organization has a form of time series representing continuous time random walk \cite{Hu16,Metz16,Metz14}. It is very likely, that the natively disordered transcription factors, in their one-dimensional search for its target on DNA, intermittent by three-dimensional flights \cite{Tafv11,Li11}, perform not the passive diffusion but the active continuous time random walk.

As mentioned earlier \cite{Kurz14}, our approach is able to explain the observation, that the single myosin II head can take several steps along the actin filament per ATP molecule hydrolyzed \cite{Kita97,Kita05}. It is likely that the mechanism of the action of the small G-proteins such as Ras, which have a common ancestor with the myosin \cite{Kull98} and an alike partly disordered structure after binding the nucleoside triphosphate \cite{Houd01,Kosz02,Arai06}, is, after a malignant transformation, similar. Of course, we cannot claim that the model considered here has something to do with the dynamics of the myosin had or the small G-proteins, but in future, a more adequate model is worth considering. The greatest weakness of the current model is the temperature independence of conformational transitions and this is also considered to be improved in the near future. 

In the healthy cells, the G-proteins play the role of signal transducers that activate various kinases, which further transmit the signal to the target. The chemotactic response of {\em Escherichia coli} to methylation of a receptor, leading to activation of the flagellar motor to move, is the relatively simple experimental model investigated in more detail \cite{Tu08}. Theory of the biochemical signal transduction in this system is the subject of intensive research \cite{Ito13,Bara13,Bara14a,Sart14,Ito15,Bo15,Hart16}. Here, the information is also stored in the form of the concentration of specific molecules like in the system considered in the present paper. It would be interesting to combine both approaches.

There are some arguments, for instance, for the kinesin motor \cite{Tani07,Bier07,Astu07,Liep09,Astu10}, the myosin V motor \cite{Nish08}, the cytoplasmic dynein motor \cite{Tsyg09,Tsyg11}, the quinol:cytochrome c oxidoreductase \cite{Swie10,Sare15}, and the very Ras molecules \cite{Iver14,Naka16} that an organization transduction with $\epsilon = 1$ is achieved in dimeric protein complexes, which are composed of two identical monomers. We have ascertained that the case of the total compensation of the entropy production by information creation is the optimum in the sense that the cost of the total process is then zero. This may occur for a binary system, whose components create and collect information alternately in such a way that the resultant information and entropy productions are zero. The cyclic, alternate behavior is important for the corresponding cost in the stationary process could not be assigned to individual components. In this respect, the corresponding models should differ essentially from those of the bipartite systems \cite{Hart14,Horo14}, and are worth a separate study. Research of this kind could help to answer the certainly interesting question: why do most protein machines operate as dimers or higher organized structures?

The main conclusion of the paper is that the free energy processing has to be distinguished from the organization processing. From the former point of view, Maxwell's demon utilizes entropy reduction for the performance of work and more precisely, for a reduction of energy losses. From the latter point of view, it can be used for other purposes, for example molecular recognition. Our answer to the question posed in the title is that the biological molecular machines can, under certain conditions, act as Maxwell's demons, but only creating information, and not performing work. This statement is based on the relationship $\epsilon < n$ between the degree of coupling $\epsilon$ and the transmission ratio $n$ shown to hold for the studied model of dynamics. There is still the need to prove the generality of this relationship.

We know that work, heat and dissipation (the entropy production) are changes of energy. But there is still controversy, the change of which physical quantity is information \cite{Parr15,Crut12}? Here, we suggest the answer that information is the change of organization, a quantity being the difference of two physical quantities $X_2$ and $X_1$ taking part in the free energy-transduction process and describing the free energy reception and delivery, respectively. In such an approach, the dual nature of entropy becomes clear as the physical quantity that connects the processes of the free energy transduction and the organization transduction.

At the end, let us take the liberty for a couple of speculation. The first and second laws of thermodynamics (\ref{eq16}) and (\ref{eq17}) are valid for any isothermal processes, whereas the suggested first and second laws of arrangement transduction (\ref{eq18}) were justified only for the stationary isothermal processes. The open problem remains the generality of these laws. One thing is certain: the necessary conditions for the information and entropy productions to be of the opposite signs, is the presence of fluctuations and the possibility of a choice. Besides the mesoscopic machines, we know three macroscopic systems sharing such properties and intriguing long. The first and the best known are the systems with critical thermodynamic fluctuations, whose organization is determined by new thermodynamic variables that survive stochastization \cite{Penr70}, referred to as the order parameters \cite{Call85} or, in various contexts, the emergent \cite{Ande72} or structural \cite{Kurz06} variables. Here, the long-living transient stage can be identified, e.g., with the nonergodic  condensation of gas through the state of fogg or the nonergodic solidification (the glass transition) of liquid \cite{Call85}. The second example are the systems displaying quantum fluctuations entangled with the environment, which organization is determined by classical variables that survive decoherence \cite{Zure03,Zure09}. And the third, most controviersial example are living systems that display a non-gradual stochastic Darwinian evolution, and their organization is determined by the survival degree of a species, long-resistant against mutations \cite{Eldr72,Bak93}.

\vspace{2pc}

\noindent
{\bf Acknowledgements.}
MK thanks Yasar Demirel and Herve Cailleau for discussing the problem in the early stages of the investigation.

\noindent
{\bf Author Contributions.}
The general concept and the theory is mainly due to MK, who also wrote the manuscript. The specification of the critical branching tree model and the numerical simulations are mainly due to PC.

\noindent
{\bf Conflict of Interest.}
None declared.


\begin{thebibliography}{00}

\expandafter\ifx\csname natexlab\endcsname\relax\def\natexlab#1{#1}\fi
\expandafter\ifx\csname bibnamefont\endcsname\relax
  \def\bibnamefont#1{#1}\fi
\expandafter\ifx\csname bibfnamefont\endcsname\relax
  \def\bibfnamefont#1{#1}\fi
\expandafter\ifx\csname citenamefont\endcsname\relax
  \def\citenamefont#1{#1}\fi
\expandafter\ifx\csname url\endcsname\relax
  \def\url#1{\texttt{#1}}\fi
\expandafter\ifx\csname urlprefix\endcsname\relax\def\urlprefix{URL }\fi
\providecommand{\bibinfo}[2]{#2}
\providecommand{\eprint}[2][]{\url{#2}}

\bibitem{Maxw71}
\bibinfo{author}{\bibfnamefont{J.~C.}~\bibnamefont{Maxwell}}, 
    \bibinfo{title}{Theory of Heat},
    \bibinfo{publisher}{Logmans,~London}, \bibinfo{year}{1871}.

\bibitem{Szil29}
\bibinfo{author}{\bibfnamefont{L.~Z.}~\bibnamefont{Szilard}}, 
    \bibinfo{journal}{Z.~ Phys.}
    \bibinfo{volume}{53} \bibinfo{year}{(1929)} \bibinfo{page}{840-857}.

\bibitem{Land61}
\bibinfo{author}{\bibfnamefont{R.}~\bibnamefont{Landauer}}, 
    \bibinfo{journal}{IBM J.~Res.~Dev.}
    \bibinfo{volume}{5} \bibinfo{year}{(1961)} \bibinfo{page}{183-191}.

\bibitem{Leef03}
\bibinfo{author}{\bibfnamefont{H.~S.}~\bibnamefont{Leef}},
\bibinfo{author}{\bibfnamefont{E.}~\bibnamefont{Rex eds.}}, 
    \bibinfo{title}{Maxwell's Demon 2: Entropy, Classical and Quantum Information, Computing},
    \bibinfo{publisher}{Institute of Physics Publishing,~Philadelphia}, \bibinfo{year}{2003}.

\bibitem{Maru09}
\bibinfo{author}{\bibfnamefont{K.}~\bibnamefont{Maruyama}},
\bibinfo{author}{\bibfnamefont{F.}~\bibnamefont{Nori}},
\bibinfo{author}{\bibfnamefont{V.}~\bibnamefont{Verdal}},
    \bibinfo{journal}{Revs.~Mod.~Phys.}
    \bibinfo{volume}{81} \bibinfo{year}{(2009)} \bibinfo{page}{1-23}.

\bibitem{Saga13}
\bibinfo{author}{\bibfnamefont{T.}~\bibnamefont{Sagawa}}, 
    \bibinfo{title}{Thermodynamics of Information Processing in Small Systems},
    \bibinfo{publisher}{Springer,~Berlin}, \bibinfo{year}{2013}.

\bibitem{Saga10}
\bibinfo{author}{\bibfnamefont{T.}~\bibnamefont{Sagawa}},
\bibinfo{author}{\bibfnamefont{M.}~\bibnamefont{Ueda}},
    \bibinfo{journal}{Phys.~Rev.~Lett.}
    \bibinfo{volume}{104} \bibinfo{year}{(2010)} \bibinfo{page}{090602}.

\bibitem{Ponm10}
\bibinfo{author}{\bibfnamefont{M.}~\bibnamefont{Ponmurugan}},
    \bibinfo{journal}{Phys.~Rev.~E}
    \bibinfo{volume}{82} \bibinfo{year}{(2010)} \bibinfo{page}{031129}.

\bibitem{Saga12}
\bibinfo{author}{\bibfnamefont{T.}~\bibnamefont{Sagawa}},
\bibinfo{author}{\bibfnamefont{M.}~\bibnamefont{Ueda}},
    \bibinfo{journal}{Phys.~Rev.~E}
    \bibinfo{volume}{85} \bibinfo{year}{(2012)} \bibinfo{page}{021104}.

\bibitem{Saga13a}
\bibinfo{author}{\bibfnamefont{T.}~\bibnamefont{Sagawa}},
\bibinfo{author}{\bibfnamefont{M.}~\bibnamefont{Ueda}},
    \bibinfo{journal}{New~J.~Phys.}
    \bibinfo{volume}{15} \bibinfo{year}{(2013)} \bibinfo{page}{125012}.

\bibitem{Hart14}
\bibinfo{author}{\bibfnamefont{B.}~\bibnamefont{Hartich}},
\bibinfo{author}{\bibfnamefont{A.~C.}~\bibnamefont{Barato}},
\bibinfo{author}{\bibfnamefont{U.}~\bibnamefont{Seifert}},
    \bibinfo{journal}{J.~Stat.~Mech.}
    \bibinfo{volume}{14} \bibinfo{year}{(2014)} \bibinfo{page}{P02016}.

\bibitem{Horo14}
\bibinfo{author}{\bibfnamefont{J.~M.}~\bibnamefont{Horowitz}},
\bibinfo{author}{\bibfnamefont{M.}~\bibnamefont{Esposito}},
    \bibinfo{journal}{Phys.~Rev.~X}
    \bibinfo{volume}{4} \bibinfo{year}{(2014)} \bibinfo{page}{031015}.

\bibitem{Mand12}
\bibinfo{author}{\bibfnamefont{D.}~\bibnamefont{Mandal}},
\bibinfo{author}{\bibfnamefont{C.}~\bibnamefont{Jarzynski}},
    \bibinfo{journal}{Proc.~Natl.~Acad.~Sci.~USA}
    \bibinfo{volume}{109} \bibinfo{year}{(2012)} \bibinfo{page}{11641-11645}.

\bibitem{Bara14}
\bibinfo{author}{\bibfnamefont{A.~C.}~\bibnamefont{Barato}},
\bibinfo{author}{\bibfnamefont{U.}~\bibnamefont{Seifert}}, 
    \bibinfo{journal}{Phys.~Rev.~Lett.}
    \bibinfo{volume}{112} \bibinfo{year}{(2014)} \bibinfo{page}{090601}.

\bibitem{Horo13}
\bibinfo{author}{\bibfnamefont{J.~M.}~\bibnamefont{Horowitz}},
\bibinfo{author}{\bibfnamefont{T.}~\bibnamefont{Sagawa}},
\bibinfo{author}{\bibfnamefont{J.~M.~R.}~\bibnamefont{Parrondo}},
    \bibinfo{journal}{Phys.~Rev.~Lett.}
    \bibinfo{volume}{111} \bibinfo{year}{(2013)} \bibinfo{page}{010602}.

\bibitem{Shir15}
\bibinfo{author}{\bibfnamefont{N.}~\bibnamefont{Shiraishi}},
\bibinfo{author}{\bibfnamefont{T.}~\bibnamefont{Sagawa}}, 
    \bibinfo{journal}{Phys.~Rev.~E}
    \bibinfo{volume}{91} \bibinfo{year}{(2015)} \bibinfo{page}{012130}.

\bibitem{Shir15a}
\bibinfo{author}{\bibfnamefont{N.}~\bibnamefont{Shiraishi}},
\bibinfo{author}{\bibfnamefont{T.}~\bibnamefont{Ito}},
\bibinfo{author}{\bibfnamefont{K.}~\bibnamefont{Kawaguchi}},
\bibinfo{author}{\bibfnamefont{T.}~\bibnamefont{Sagawa}}, 
    \bibinfo{journal}{New~J.~Phys.}
    \bibinfo{volume}{17} \bibinfo{year}{(2015)} \bibinfo{page}{045012}.

\bibitem{Shir16}
\bibinfo{author}{\bibfnamefont{N.}~\bibnamefont{Shiraishi}},
\bibinfo{author}{\bibfnamefont{T.}~\bibnamefont{Matsumoto}},
\bibinfo{author}{\bibfnamefont{T.}~\bibnamefont{Sagawa}}, 
    \bibinfo{journal}{New~J.~Phys.}
    \bibinfo{volume}{18} \bibinfo{year}{(2016)} \bibinfo{page}{013044}.

\bibitem{Deff13}
\bibinfo{author}{\bibfnamefont{S.}~\bibnamefont{Deffner}},
\bibinfo{author}{\bibfnamefont{C.}~\bibnamefont{Jarzynski}},
    \bibinfo{journal}{Phys.~Rev.~X}
    \bibinfo{volume}{3} \bibinfo{year}{(2013)} \bibinfo{page}{041003}.

\bibitem{Smol12}
\bibinfo{author}{\bibfnamefont{M.}~\bibnamefont{Smoluchowski}},
    \bibinfo{journal}{Phys.~Z.}
    \bibinfo{volume}{13} \bibinfo{year}{(1912)} \bibinfo{page}{1069-1080}.

\bibitem{Feyn63}
\bibinfo{author}{\bibfnamefont{R.~P.}~\bibnamefont{Feynman}},
\bibinfo{author}{\bibfnamefont{R.~B.}~\bibnamefont{Leighton}},
\bibinfo{author}{\bibfnamefont{M.}~\bibnamefont{Sands}}, 
    \bibinfo{title}{The Feynman Lectures on Physics, vol. 1, chap. 46},
    \bibinfo{publisher}{Addison-Wesley,~Reading}, \bibinfo{year}{1963}.

\bibitem{Jarz99}
\bibinfo{author}{\bibfnamefont{C.}~\bibnamefont{Jarzynski}},
\bibinfo{author}{\bibfnamefont{O.}~\bibnamefont{Mazonka}},
    \bibinfo{journal}{Phys.~Rev.~E}
    \bibinfo{volume}{59} \bibinfo{year}{(1999)} \bibinfo{page}{6448-6459}.

\bibitem{Huxl57}
\bibinfo{author}{\bibfnamefont{A.~F.}~\bibnamefont{Haxley}},
    \bibinfo{journal}{Prog.~ Biophys.~Biophys.~Chem.}
    \bibinfo{volume}{7} \bibinfo{year}{(1957)} \bibinfo{page}{255-318}.

\bibitem{Howa01}
\bibinfo{author}{\bibfnamefont{J.}~\bibnamefont{Howard}},
    \bibinfo{title}{Mechanics of Motor Proteins and the Cytoskeleton},
    \bibinfo{publisher}{Sinauer, Sunderland}, \bibinfo{year}{2001}.

\bibitem{Howa10}
\bibinfo{author}{\bibfnamefont{J.}~\bibnamefont{Howard}},
    \bibinfo{title}{Motor proteins as nanomachines: the roles of thermal fluctuations in generating force and motion},
    \bibinfo{journal}{in Biological Physics - Poincare Seminar 2009},
    \bibinfo{editors}{B.~Duplainer,~V.~Rivasseau, eds.}, \bibinfo{page}{47-60},
    \bibinfo{publisher}{Springer,~Basel}, \bibinfo{year}{2003}.

\bibitem{Kurz98}
\bibinfo{author}{\bibfnamefont{M.}~\bibnamefont{Kurzynski}},
    \bibinfo{journal}{Prog.~Biophys.~Molec.~Biol.}
    \bibinfo{volume}{69} \bibinfo{year}{(1998)} \bibinfo{page}{23-82}.

\bibitem{Fish99}
\bibinfo{author}{\bibfnamefont{M.~A.}~\bibnamefont{Fisher}},
\bibinfo{author}{\bibfnamefont{A.~B.}~\bibnamefont{Kolomeisky}},
    \bibinfo{journal}{Proc.~Natl.~Acad.~Sci.~USA}
    \bibinfo{volume}{96} \bibinfo{year}{(1999)} \bibinfo{page}{6597-6602}.

\bibitem{Kolo07}
\bibinfo{author}{\bibfnamefont{A.~B.}~\bibnamefont{Kolomeisky}},
\bibinfo{author}{\bibfnamefont{M.~A.}~\bibnamefont{Fisher}},
    \bibinfo{journal}{Annu.~Rev.~Phys.~Chem.}
    \bibinfo{volume}{58} \bibinfo{year}{(2007)} \bibinfo{page}{675-695}.

\bibitem{Astu99}
\bibinfo{author}{\bibfnamefont{R.~D.}~\bibnamefont{Astumian}},
\bibinfo{author}{\bibfnamefont{I.}~\bibnamefont{Derenyi}},
    \bibinfo{journal}{Biophys.~J.}
    \bibinfo{volume}{77} \bibinfo{year}{(1999)} \bibinfo{page}{993-1002}.

\bibitem{Bust01}
\bibinfo{author}{\bibfnamefont{C.}~\bibnamefont{Bustamante}},
\bibinfo{author}{\bibfnamefont{D.}~\bibnamefont{Keller}},
\bibinfo{author}{\bibfnamefont{G.}~\bibnamefont{Oster}},
    \bibinfo{journal}{Acc.~Chem.~Res.}
    \bibinfo{volume}{34} \bibinfo{year}{(2001)} \bibinfo{page}{412-420}.

\bibitem{Lipo00}
\bibinfo{author}{\bibfnamefont{R.}~\bibnamefont{Lipowsky}},
    \bibinfo{journal}{Phys.~Rev.~Lett.}
    \bibinfo{volume}{85} \bibinfo{year}{(2000)} \bibinfo{page}{4401-4406}.

\bibitem{Lipo09}
\bibinfo{author}{\bibfnamefont{R.}~\bibnamefont{Lipowsky}},
\bibinfo{author}{\bibfnamefont{S.}~\bibnamefont{Liepelt}},
\bibinfo{author}{\bibfnamefont{A.}~\bibnamefont{Valleriani}},
    \bibinfo{journal}{J.~Stat.~Phys.}
    \bibinfo{volume}{135} \bibinfo{year}{(2009)} \bibinfo{page}{951-975}.

\bibitem{Kurz03}
\bibinfo{author}{\bibfnamefont{M.}~\bibnamefont{Kurzynski}},
\bibinfo{author}{\bibfnamefont{P.}~\bibnamefont{Chelminiak}},
    \bibinfo{journal}{J.~Stat.~Phys.}
    \bibinfo{volume}{110} \bibinfo{year}{(2003)} \bibinfo{page}{137-181}.

\bibitem{Kurz14}
\bibinfo{author}{\bibfnamefont{M.}~\bibnamefont{Kurzynski}},
\bibinfo{author}{\bibfnamefont{M.}~\bibnamefont{Torchala}},
\bibinfo{author}{\bibfnamefont{P.}~\bibnamefont{Chelminiak}},
    \bibinfo{journal}{Phys.~Rev.~E}
    \bibinfo{volume}{89} \bibinfo{year}{(2014)} \bibinfo{page}{012722}.

\bibitem{Jarz11}
\bibinfo{author}{\bibfnamefont{C.}~\bibnamefont{Jarzynski}},
    \bibinfo{journal}{Annu.~Rev.~Condens.~Matter~Phys.}
    \bibinfo{volume}{2} \bibinfo{year}{(2011)} \bibinfo{page}{329-351}.

\bibitem{Seif12}
\bibinfo{author}{\bibfnamefont{U.}~\bibnamefont{Seifert}},
    \bibinfo{journal}{Rep.~Prog.~Phys.}
    \bibinfo{volume}{75} \bibinfo{year}{(2012)} \bibinfo{page}{126001(58pp)}.

\bibitem{Parr15}
\bibinfo{author}{\bibfnamefont{J.~M.~R.}~\bibnamefont{Parrondo}},
\bibinfo{author}{\bibfnamefont{J.~M.}~\bibnamefont{Horowitz}},
\bibinfo{author}{\bibfnamefont{T.}~\bibnamefont{Sagawa}},
    \bibinfo{journal}{Nature~Phys.}
    \bibinfo{volume}{11} \bibinfo{year}{(2015)} \bibinfo{page}{131-139}.

\bibitem{Toya10}
\bibinfo{author}{\bibfnamefont{S.}~\bibnamefont{Toyabe}},
\bibinfo{author}{\bibfnamefont{T.}~\bibnamefont{Sagawa}},
\bibinfo{author}{\bibfnamefont{M.}~\bibnamefont{Ueda}},
\bibinfo{author}{\bibfnamefont{E.}~\bibnamefont{Muneyuki}},
\bibinfo{author}{\bibfnamefont{M.}~\bibnamefont{Sano}},
    \bibinfo{journal}{Nature~Phys.}
    \bibinfo{volume}{6} \bibinfo{year}{(2010)} \bibinfo{page}{988-992}.

\bibitem{Beru12}
\bibinfo{author}{\bibfnamefont{A.}~\bibnamefont{B\'erut}},
\bibinfo{author}{\bibfnamefont{A.}~\bibnamefont{Arakelyan}},
\bibinfo{author}{\bibfnamefont{A.}~\bibnamefont{Petrosyan}},
\bibinfo{author}{\bibfnamefont{S.}~\bibnamefont{Ciliberto}},
\bibinfo{author}{\bibfnamefont{R.}~\bibnamefont{Dillenschneider}},
\bibinfo{author}{\bibfnamefont{E.}~\bibnamefont{Lutz}},
    \bibinfo{journal}{Nature}
    \bibinfo{volume}{483} \bibinfo{year}{(2012)} \bibinfo{page}{187-190}.

\bibitem{Kosk14}
\bibinfo{author}{\bibfnamefont{J.~V.}~\bibnamefont{Koski}},
\bibinfo{author}{\bibfnamefont{Y.~F.}~\bibnamefont{Maisi}},
\bibinfo{author}{\bibfnamefont{T.}~\bibnamefont{Sagawa}},
\bibinfo{author}{\bibfnamefont{E.}~\bibnamefont{Pekola}},
    \bibinfo{journal}{Phys.~Rev.~Lett.}
    \bibinfo{volume}{113} \bibinfo{year}{(2014)} \bibinfo{page}{030601}.

\bibitem{Rold14}
\bibinfo{author}{\bibfnamefont{\'E.}~\bibnamefont{Rold\'an}},
\bibinfo{author}{\bibfnamefont{I.~A.}~\bibnamefont{Martinez}},
\bibinfo{author}{\bibfnamefont{J.~M.~R}~\bibnamefont{Parrondo}},
\bibinfo{author}{\bibfnamefont{D.}~\bibnamefont{Petrov}},
    \bibinfo{journal}{Nature~Phys.}
    \bibinfo{volume}{10} \bibinfo{year}{(2014)} \bibinfo{page}{457-461}.

\bibitem{Roze10}
\bibinfo{author}{\bibfnamefont{H.~D.}~\bibnamefont{Rozenfeld}},
\bibinfo{author}{\bibfnamefont{C.}~\bibnamefont{Song}},
\bibinfo{author}{\bibfnamefont{H.~A.}~\bibnamefont{Makse}},
    \bibinfo{journal}{Phys.~Rev.~Lett.}
    \bibinfo{volume}{104} \bibinfo{year}{(2010)} \bibinfo{page}{025701}.

\bibitem{Esco12}
\bibinfo{author}{\bibfnamefont{F.}~\bibnamefont{Escolano}},
\bibinfo{author}{\bibfnamefont{E.~R.}~\bibnamefont{Handcock}},
\bibinfo{author}{\bibfnamefont{M.~A.}~\bibnamefont{Lozano}},
    \bibinfo{journal}{Phys.~Rev.~E}
    \bibinfo{volume}{85} \bibinfo{year}{(2012)} \bibinfo{page}{036206}.

\bibitem{Kurz06}
\bibinfo{author}{\bibfnamefont{M.}~\bibnamefont{Kurzynski}},
    \bibinfo{title}{The Thermodynamic Machinery of Life},
    \bibinfo{publisher}{Springer,~Berlin}, \bibinfo{year}{2006}.

\bibitem{Call85}
\bibinfo{author}{\bibfnamefont{H.~B.}~\bibnamefont{Callen}},
    \bibinfo{title}{Thermodynamics and an Introduction to Thermostatistics 2nd edn.},
    \bibinfo{publisher}{Wiley,~New York}, \bibinfo{year}{1989}.

\bibitem{Hill89}
\bibinfo{author}{\bibfnamefont{T.~L.}~\bibnamefont{Hill}},
    \bibinfo{title}{Free Energy Transduction and Biochemical Cycle Kinetics},
    \bibinfo{publisher}{Springer,~New York}, \bibinfo{year}{1989}.

\bibitem{Kita97}
\bibinfo{author}{\bibfnamefont{K.}~\bibnamefont{Kitamura}},
\bibinfo{author}{\bibfnamefont{M.}~\bibnamefont{Tokunaga}},
\bibinfo{author}{\bibfnamefont{A.~H.}~\bibnamefont{Iwane}},
\bibinfo{author}{\bibfnamefont{T.}~\bibnamefont{Yanagida}},
    \bibinfo{journal}{Nature}
    \bibinfo{volume}{397} \bibinfo{year}{(1997)} \bibinfo{page}{129-134}.

\bibitem{Kita05}
\bibinfo{author}{\bibfnamefont{K.}~\bibnamefont{Kitamura}},
\bibinfo{author}{\bibfnamefont{M.}~\bibnamefont{Tokunaga}},
\bibinfo{author}{\bibfnamefont{S.}~\bibnamefont{Esaki}},
\bibinfo{author}{\bibfnamefont{A.~H.}~\bibnamefont{Iwane}},
\bibinfo{author}{\bibfnamefont{T.}~\bibnamefont{Yanagida}},
    \bibinfo{journal}{Biophysics}
    \bibinfo{volume}{1} \bibinfo{year}{(2005)} \bibinfo{page}{1-19}.

\bibitem{Andr07}
\bibinfo{author}{\bibfnamefont{D.}~\bibnamefont{Andrieux}},
\bibinfo{author}{\bibfnamefont{P.}~\bibnamefont{Gaspard}},
    \bibinfo{journal}{J.~Stat.~Phys.}
    \bibinfo{volume}{127} \bibinfo{year}{(2007)} \bibinfo{page}{107}.

\bibitem{Gasp13}
\bibinfo{author}{\bibfnamefont{P.}~\bibnamefont{Gaspard}},
    \bibinfo{journal}{New~J.~Phys.}
    \bibinfo{volume}{15} \bibinfo{year}{(2013)} \bibinfo{page}{115014}.

\bibitem{Jarz97}
\bibinfo{author}{\bibfnamefont{C.}~\bibnamefont{Jarzynski}},
    \bibinfo{journal}{Phys.~Rev.~Lett.}
    \bibinfo{volume}{78} \bibinfo{year}{(1997)} \bibinfo{page}{2690-2693}.

\bibitem{Mehl12}
\bibinfo{author}{\bibfnamefont{J.}~\bibnamefont{Mehl}},
\bibinfo{author}{\bibfnamefont{B.}~\bibnamefont{Lander}},
\bibinfo{author}{\bibfnamefont{C.}~\bibnamefont{Bechinger}},
\bibinfo{author}{\bibfnamefont{V.}~\bibnamefont{Blickle}},
\bibinfo{author}{\bibfnamefont{U.}~\bibnamefont{Seifert}},
    \bibinfo{journal}{Phys.~Rev.~Lett.}
    \bibinfo{volume}{108} \bibinfo{year}{(2012)} \bibinfo{page}{220601}.

\bibitem{Borr15}
\bibinfo{author}{\bibfnamefont{M.}~\bibnamefont{Borrelli}},
\bibinfo{author}{\bibfnamefont{J.~V.}~\bibnamefont{Koski}},
\bibinfo{author}{\bibfnamefont{S.}~\bibnamefont{Maniscalco}},
\bibinfo{author}{\bibfnamefont{J.~P.}~\bibnamefont{Pekola}},
    \bibinfo{journal}{Phys.~Rev.~E}
    \bibinfo{volume}{91} \bibinfo{year}{(2015)} \bibinfo{page}{012145}.

\bibitem{Henz07}
\bibinfo{author}{\bibfnamefont{K.}~\bibnamefont{Henzler-Wildman}},
\bibinfo{author}{\bibfnamefont{D.}~\bibnamefont{Kern}},
    \bibinfo{journal}{Nature}
    \bibinfo{volume}{450} \bibinfo{year}{(2007)} \bibinfo{page}{964-972}.

\bibitem{Uver10}
\bibinfo{author}{\bibfnamefont{V.~N.}~\bibnamefont{Uversky}},
\bibinfo{author}{\bibfnamefont{A.~K.}~\bibnamefont{Dunker}},
    \bibinfo{journal}{Biochim.~Biophys.~Acta}
    \bibinfo{volume}{1804} \bibinfo{year}{(2010)} \bibinfo{page}{1231-1263}.

\bibitem{Chou11}
\bibinfo{author}{\bibfnamefont{D.}~\bibnamefont{Chouard}},
    \bibinfo{journal}{Nature}
    \bibinfo{volume}{471} \bibinfo{year}{(2011)} \bibinfo{page}{151-153}.

\bibitem{Chod14}
\bibinfo{author}{\bibfnamefont{J.~D.}~\bibnamefont{Chodera}},
\bibinfo{author}{\bibfnamefont{F.}~\bibnamefont{Noe}},
    \bibinfo{journal}{Curr.~Opin.~Struct.~Biol.}
    \bibinfo{volume}{25} \bibinfo{year}{(2014)} \bibinfo{page}{135-144}.

\bibitem{Shuk15}
\bibinfo{author}{\bibfnamefont{D.}~\bibnamefont{Shukla}},
\bibinfo{author}{\bibfnamefont{C.~X.}~\bibnamefont{Hernandez}},
\bibinfo{author}{\bibfnamefont{J.~K.}~\bibnamefont{Weber}},
\bibinfo{author}{\bibfnamefont{V.~S.}~\bibnamefont{Pande}},
    \bibinfo{journal}{Acc.~Chem.~Res.}
    \bibinfo{volume}{48} \bibinfo{year}{(2015)} \bibinfo{page}{414-422}.

\bibitem{Aust75}
\bibinfo{author}{\bibfnamefont{R.~H.}~\bibnamefont{Austin}},
\bibinfo{author}{\bibfnamefont{K.~W.}~\bibnamefont{Beeson}},
\bibinfo{author}{\bibfnamefont{L.}~\bibnamefont{Eisenstein}},
\bibinfo{author}{\bibfnamefont{H.}~\bibnamefont{Frauenfelder}},
\bibinfo{author}{\bibfnamefont{I.~C.}~\bibnamefont{Gunsalus}},
    \bibinfo{journal}{Biochemistry}
    \bibinfo{volume}{14} \bibinfo{year}{(1975)} \bibinfo{page}{5355-5373}.

\bibitem{Frau10}
\bibinfo{author}{\bibfnamefont{H.}~\bibnamefont{Frauenfelder}},
    \bibinfo{title}{The Physics of Proteins: An Introduction to Biological Physics and Molecular Biophysics},
    \bibinfo{publisher}{Springer,~Berlin}, \bibinfo{year}{2010}.

\bibitem{Ishi01}
\bibinfo{author}{\bibfnamefont{A.}~\bibnamefont{Ishijima}},
\bibinfo{author}{\bibfnamefont{T.}~\bibnamefont{Yanagida}},
    \bibinfo{journal}{Trends Biochem.~Sci.}
    \bibinfo{volume}{26} \bibinfo{year}{(2001)} \bibinfo{page}{438-444}.

\bibitem{Yana08}
\bibinfo{author}{\bibfnamefont{T.}~\bibnamefont{Yanagida}},
\bibinfo{author}{\bibfnamefont{E.}~\bibnamefont{Ishii, eds.}},
    \bibinfo{title}{Single Molecule Dynamics in Life Science},
    \bibinfo{publisher}{Wiley-VCH, Weinheim}, \bibinfo{year}{2008}.

\bibitem{Kita98}
\bibinfo{author}{\bibfnamefont{A.}~\bibnamefont{Kitao}},
\bibinfo{author}{\bibfnamefont{S.}~\bibnamefont{Hayward}},
\bibinfo{author}{\bibfnamefont{N.}~\bibnamefont{Go}},
    \bibinfo{journal}{Proteins}
    \bibinfo{volume}{33} \bibinfo{year}{(1998)} \bibinfo{page}{496-517}.

\bibitem{Karp02}
\bibinfo{author}{\bibfnamefont{M.}~\bibnamefont{Karplus}},
\bibinfo{author}{\bibfnamefont{J.~A.}~\bibnamefont{McCammon}},
    \bibinfo{journal}{Nature Struct.~ Biol.}
    \bibinfo{volume}{9} \bibinfo{year}{(2002)} \bibinfo{page}{646-652}.

\bibitem{Yang09}
\bibinfo{author}{\bibfnamefont{S.}~\bibnamefont{Yang}},
\bibinfo{author}{\bibfnamefont{N.~K.}~\bibnamefont{Banavali}},
\bibinfo{author}{\bibfnamefont{B.}~\bibnamefont{Roux}},
    \bibinfo{journal}{Proc.~Natl.~Acad.~Sci.~USA.}
    \bibinfo{volume}{106} \bibinfo{year}{(2009)} \bibinfo{page}{3776-3781}.

\bibitem{Hu16}
\bibinfo{author}{\bibfnamefont{X.}~\bibnamefont{Hu}},
\bibinfo{author}{\bibfnamefont{L.}~\bibnamefont{Hong}},
\bibinfo{author}{\bibfnamefont{M.~D.}~\bibnamefont{Smith}},
\bibinfo{author}{\bibfnamefont{T.}~\bibnamefont{Neusius}},
\bibinfo{author}{\bibfnamefont{X.}~\bibnamefont{Cheng}},
\bibinfo{author}{\bibfnamefont{J.~C.}~\bibnamefont{Smith}},
    \bibinfo{journal}{Nature Phys.}
    \bibinfo{volume}{12} \bibinfo{year}{(2016)} \bibinfo{page}{171-174}.

\bibitem{Zwan90}
\bibinfo{author}{\bibfnamefont{R.}~\bibnamefont{Zwanzig}},
    \bibinfo{journal}{Acc.~Chem.~Res.}
    \bibinfo{volume}{23} \bibinfo{year}{(1990)} \bibinfo{page}{148-152}.

\bibitem{Lu98}
\bibinfo{author}{\bibfnamefont{H.~P.}~\bibnamefont{Lu}},
\bibinfo{author}{\bibfnamefont{L.}~\bibnamefont{Xun}},
\bibinfo{author}{\bibfnamefont{X.~S.}~\bibnamefont{Xie}},
    \bibinfo{journal}{Science}
    \bibinfo{volume}{282} \bibinfo{year}{(1998)} \bibinfo{page}{1877-1882}.

\bibitem{Engl06}
\bibinfo{author}{\bibfnamefont{B.~P.}~\bibnamefont{English}},
\bibinfo{author}{\bibfnamefont{W.}~\bibnamefont{Min}},
\bibinfo{author}{\bibfnamefont{A.~M.}~\bibnamefont{van Oijen}},
\bibinfo{author}{\bibfnamefont{K.~T.}~\bibnamefont{Lee}},
\bibinfo{author}{\bibfnamefont{G.}~\bibnamefont{Luo}},
\bibinfo{author}{\bibfnamefont{H.}~\bibnamefont{Sun}},
    \bibinfo{journal}{Nature Chem.~Biol.}
    \bibinfo{volume}{2} \bibinfo{year}{(2006)} \bibinfo{page}{87-94}.

\bibitem{Kurz08}
\bibinfo{author}{\bibfnamefont{M.}~\bibnamefont{Kurzynski}},
\bibinfo{journal}{Cell.~Mol.~Biol.~Lett.}
    \bibinfo{volume}{13} \bibinfo{year}{(2008)} \bibinfo{page}{502-513}.

\bibitem{Xie13}
\bibinfo{author}{\bibfnamefont{X.~S.}~\bibnamefont{Xie}},
\bibinfo{journal}{Science}
    \bibinfo{volume}{342} \bibinfo{year}{(2013)} \bibinfo{page}{1457-1459}.

\bibitem{Bak96}
\bibinfo{author}{\bibfnamefont{P.}~\bibnamefont{Bak}},
    \bibinfo{title}{How Nature Works},
    \bibinfo{publisher}{Springer,~New York}, \bibinfo{year}{1996}.

\bibitem{Snepp05}
\bibinfo{author}{\bibfnamefont{K.}~\bibnamefont{Sneppen}},
\bibinfo{author}{\bibfnamefont{G.}~\bibnamefont{Zocchi}},
    \bibinfo{title}{Physics in Molecular Biology},
    \bibinfo{publisher}{Cambridge University Press,~New York}, \bibinfo{year}{2005}.

\bibitem{Goh06}
\bibinfo{author}{\bibfnamefont{K.~I.}~\bibnamefont{Goh}},
\bibinfo{author}{\bibfnamefont{G.}~\bibnamefont{Salvi}},
\bibinfo{author}{\bibfnamefont{B.}~\bibnamefont{Kahng}},
\bibinfo{author}{\bibfnamefont{D.}~\bibnamefont{Kim}},
\bibinfo{journal}{Phys.~Rev.~Lett.}
    \bibinfo{volume}{96} \bibinfo{year}{(2006)} \bibinfo{page}{018701}.

\bibitem{Chel17}
\bibinfo{author}{\bibfnamefont{P.}~\bibnamefont{Chelminiak}},
\bibinfo{author}{\bibfnamefont{M.}~\bibnamefont{Kurzynski}},
\bibinfo{journal}{Physica A}
    \bibinfo{volume}{468} \bibinfo{year}{(2017)} \bibinfo{page}{540-551}.

\bibitem{Kurz14a}
\bibinfo{author}{\bibfnamefont{M.}~\bibnamefont{Kurzynski}},
\bibinfo{author}{\bibfnamefont{P.}~\bibnamefont{Chelminiak}},
\bibinfo{journal}{Entropy}
    \bibinfo{volume}{16} \bibinfo{year}{(2014)} \bibinfo{page}{1969-1982}.

\bibitem{Palm82}
\bibinfo{author}{\bibfnamefont{R.~G.}~\bibnamefont{Palmer}},
\bibinfo{journal}{Adv.~Phys.}
    \bibinfo{volume}{31} \bibinfo{year}{(1982)} \bibinfo{page}{669-783}.

\bibitem{Metz16}
\bibinfo{author}{\bibfnamefont{R.}~\bibnamefont{Metzler}},
\bibinfo{journal}{Nature Phys.}
    \bibinfo{volume}{12} \bibinfo{year}{(2016)} \bibinfo{page}{113-114}.

\bibitem{Metz14}
\bibinfo{author}{\bibfnamefont{R.}~\bibnamefont{Metzler}},
\bibinfo{author}{\bibfnamefont{J.-H.}~\bibnamefont{Jeon}},
\bibinfo{author}{\bibfnamefont{A.~G.}~\bibnamefont{Cherstvy}},
\bibinfo{author}{\bibfnamefont{E.}~\bibnamefont{Barkai}},
\bibinfo{journal}{Phys.~Chem.~Chem.~Phys.}
    \bibinfo{volume}{16} \bibinfo{year}{(2014)} \bibinfo{page}{24128-24164}.

\bibitem{Tafv11}
\bibinfo{author}{\bibfnamefont{A.}~\bibnamefont{Tafvizi}},
\bibinfo{author}{\bibfnamefont{F.}~\bibnamefont{Huang}},
\bibinfo{author}{\bibfnamefont{A.~R.}~\bibnamefont{Ferhst}},
\bibinfo{author}{\bibfnamefont{L.~A.}~\bibnamefont{Mirny}},
\bibinfo{author}{\bibfnamefont{A.~M.~A.}~\bibnamefont{van Oijen}},
\bibinfo{journal}{Proc.~Natl.~Acad.~Sci.~USA}
    \bibinfo{volume}{108} \bibinfo{year}{(2011)} \bibinfo{page}{263-268}.

\bibitem{Li11}
\bibinfo{author}{\bibfnamefont{G.-W}~\bibnamefont{Li}},
\bibinfo{author}{\bibfnamefont{X.~S.}~\bibnamefont{Xie}},
\bibinfo{journal}{Nature}
    \bibinfo{volume}{475} \bibinfo{year}{(2011)} \bibinfo{page}{308-315}.

\bibitem{Kull98}
\bibinfo{author}{\bibfnamefont{F.~J.}~\bibnamefont{Kull}},
\bibinfo{author}{\bibfnamefont{R.~D.}~\bibnamefont{Vale}},
\bibinfo{author}{\bibfnamefont{R.~J.}~\bibnamefont{Fletterick}},
\bibinfo{journal}{J.~Muscle Res.~Cell Motil.}
    \bibinfo{volume}{19} \bibinfo{year}{(1998)} \bibinfo{page}{877-886}.

\bibitem{Houd01}
\bibinfo{author}{\bibfnamefont{A.}~\bibnamefont{Houdusse}},
\bibinfo{author}{\bibfnamefont{H.~L.}~\bibnamefont{Sweeney}},
\bibinfo{journal}{Curr.~Res.~Struct.~Biology.}
    \bibinfo{volume}{11} \bibinfo{year}{(2001)} \bibinfo{page}{182-194}.

\bibitem{Kosz02}
\bibinfo{author}{\bibfnamefont{I.}~\bibnamefont{Kosztin}},
\bibinfo{author}{\bibfnamefont{R.}~\bibnamefont{Bruinsma}},
\bibinfo{author}{\bibfnamefont{P.}~\bibnamefont{O'Lague}},
\bibinfo{author}{\bibfnamefont{K.}~\bibnamefont{Schulten}},
\bibinfo{journal}{Proc.~Natl.~Acad.~Sci.~USA}
    \bibinfo{volume}{99} \bibinfo{year}{(2002)} \bibinfo{page}{3575-3580}.

\bibitem{Arai06}
\bibinfo{author}{\bibfnamefont{Y.}~\bibnamefont{Arai}},
\bibinfo{author}{\bibfnamefont{A.~H.}~\bibnamefont{Iwane}},
\bibinfo{author}{\bibfnamefont{T.}~\bibnamefont{Wazawa}},
\bibinfo{author}{\bibfnamefont{T.}~\bibnamefont{Yokota}},
\bibinfo{author}{\bibfnamefont{Y.}~\bibnamefont{Ishii}},
\bibinfo{author}{\bibfnamefont{T.}~\bibnamefont{Kataoka}},
\bibinfo{author}{\bibfnamefont{T.}~\bibnamefont{Yanagida}},
\bibinfo{journal}{Biochem.~Biophys.~Res.~Commun.}
    \bibinfo{volume}{343} \bibinfo{year}{(2006)} \bibinfo{page}{809-815}.

\bibitem{Tu08}
\bibinfo{author}{\bibfnamefont{Y.}~\bibnamefont{Tu}},
\bibinfo{author}{\bibfnamefont{T.~S.}~\bibnamefont{Shimizu}},
\bibinfo{author}{\bibfnamefont{H.~C.}~\bibnamefont{Berg}},
\bibinfo{journal}{Proc.~Natl.~Acad.~Sci.~USA}
    \bibinfo{volume}{105} \bibinfo{year}{(2008)} \bibinfo{page}{14855-14860}.

\bibitem{Ito13}
\bibinfo{author}{\bibfnamefont{S.}~\bibnamefont{Ito}},
\bibinfo{author}{\bibfnamefont{T.}~\bibnamefont{Sagawa}},
\bibinfo{journal}{Phys.~Rev.~Lett.}
    \bibinfo{volume}{111} \bibinfo{year}{(2013)} \bibinfo{page}{180603}.

\bibitem{Bara13}
\bibinfo{author}{\bibfnamefont{A.~C.}~\bibnamefont{Barato}},
\bibinfo{author}{\bibfnamefont{D.}~\bibnamefont{Hartrich}},
\bibinfo{author}{\bibfnamefont{U.}~\bibnamefont{Seifert}},
\bibinfo{journal}{Phys.~Rev.~E}
    \bibinfo{volume}{87} \bibinfo{year}{(2013)} \bibinfo{page}{042104}.

\bibitem{Bara14a}
\bibinfo{author}{\bibfnamefont{A.~C.}~\bibnamefont{Barato}},
\bibinfo{author}{\bibfnamefont{D.}~\bibnamefont{Hartrich}},
\bibinfo{author}{\bibfnamefont{U.}~\bibnamefont{Seifert}},
\bibinfo{journal}{New J.~Phys.}
    \bibinfo{volume}{16} \bibinfo{year}{(2014)} \bibinfo{page}{103024}.

\bibitem{Sart14}
\bibinfo{author}{\bibfnamefont{P.}~\bibnamefont{Sartori}},
\bibinfo{author}{\bibfnamefont{L.}~\bibnamefont{Granger}},
\bibinfo{author}{\bibfnamefont{C.~F.}~\bibnamefont{Lee}},
\bibinfo{author}{\bibfnamefont{J.~M.}~\bibnamefont{Horowitz}},
\bibinfo{journal}{PLOS Comput.~Biol.}
    \bibinfo{volume}{10} \bibinfo{year}{(2014)} \bibinfo{page}{e1003974}.

\bibitem{Ito15}
\bibinfo{author}{\bibfnamefont{S.}~\bibnamefont{Ito}},
\bibinfo{author}{\bibfnamefont{T.}~\bibnamefont{Sagawa}},
\bibinfo{journal}{Nature Commun.}
    \bibinfo{volume}{6} \bibinfo{year}{(2015)} \bibinfo{page}{2-6}.

\bibitem{Bo15}
\bibinfo{author}{\bibfnamefont{S.}~\bibnamefont{Bo}},
\bibinfo{author}{\bibfnamefont{M.}~\bibnamefont{Del Giudice}},
\bibinfo{author}{\bibfnamefont{A.}~\bibnamefont{Celani}},
\bibinfo{journal}{J.~Stat.~Mech.}
    \bibinfo{year}{(2015)} \bibinfo{page}{P01014}.

\bibitem{Hart16}
\bibinfo{author}{\bibfnamefont{D.}~\bibnamefont{Hartrich}},
\bibinfo{author}{\bibfnamefont{A.~C.}~\bibnamefont{Barato}},
\bibinfo{author}{\bibfnamefont{U.}~\bibnamefont{Seifert}},
\bibinfo{journal}{Phys.~Rev.~E}
    \bibinfo{volume}{93} \bibinfo{year}{(2016)} \bibinfo{page}{022116}.

\bibitem{Tani07}
\bibinfo{author}{\bibfnamefont{Y.}~\bibnamefont{Taniguchi}},
\bibinfo{author}{\bibfnamefont{P.}~\bibnamefont{Karagiannis}},
\bibinfo{author}{\bibfnamefont{M.}~\bibnamefont{Nishiyama}},
\bibinfo{author}{\bibfnamefont{T.}~\bibnamefont{Yanagida}},
\bibinfo{journal}{BioSystems}
    \bibinfo{volume}{88} \bibinfo{year}{(2007)} \bibinfo{page}{283-292}.

\bibitem{Bier07}
\bibinfo{author}{\bibfnamefont{M.}~\bibnamefont{Bier}},
\bibinfo{journal}{BioSystems}
    \bibinfo{volume}{88} \bibinfo{year}{(2007)} \bibinfo{page}{301-307}.

\bibitem{Astu07}
\bibinfo{author}{\bibfnamefont{R.~D.}~\bibnamefont{Astumian}},
\bibinfo{journal}{BioSystems}
    \bibinfo{volume}{93} \bibinfo{year}{(2007)} \bibinfo{page}{8-15}.

\bibitem{Liep09}
\bibinfo{author}{\bibfnamefont{S.}~\bibnamefont{Liepelt}},
\bibinfo{author}{\bibfnamefont{R.}~\bibnamefont{Lipowsky}},
\bibinfo{journal}{Phys.~Rev.~E}
    \bibinfo{volume}{79} \bibinfo{year}{(2009)} \bibinfo{page}{011917}.

\bibitem{Astu10}
\bibinfo{author}{\bibfnamefont{R.~D.}~\bibnamefont{Astumian}},
\bibinfo{journal}{Biophys.~J.}
    \bibinfo{volume}{98} \bibinfo{year}{(2010)} \bibinfo{page}{2401-2409}.

\bibitem{Nish08}
\bibinfo{author}{\bibfnamefont{M.}~\bibnamefont{Nishikawa}},
\bibinfo{author}{\bibfnamefont{H.}~\bibnamefont{Takai}},
\bibinfo{author}{\bibfnamefont{T.}~\bibnamefont{Shibata}},
\bibinfo{author}{\bibfnamefont{A.~F.}~\bibnamefont{Ivane}},
\bibinfo{author}{\bibfnamefont{T.}~\bibnamefont{Yanagida}},
\bibinfo{journal}{Phys.~Rev.~Lett.}
    \bibinfo{volume}{101} \bibinfo{year}{(2008)} \bibinfo{page}{12103}.

\bibitem{Tsyg09}
\bibinfo{author}{\bibfnamefont{D.}~\bibnamefont{Tsygankov}},
\bibinfo{author}{\bibfnamefont{A.~W.~R.}~\bibnamefont{Serohijos}},
\bibinfo{author}{\bibfnamefont{N.~V.}~\bibnamefont{Dokholyan}},
\bibinfo{author}{\bibfnamefont{T.~C.}~\bibnamefont{Elston}},
\bibinfo{journal}{J.~Chem.~Phys.}
    \bibinfo{volume}{130} \bibinfo{year}{(2009)} \bibinfo{page}{025101}.

\bibitem{Tsyg11}
\bibinfo{author}{\bibfnamefont{D.}~\bibnamefont{Tsygankov}},
\bibinfo{author}{\bibfnamefont{A.~W.~R.}~\bibnamefont{Serohijos}},
\bibinfo{author}{\bibfnamefont{N.~V.}~\bibnamefont{Dokholyan}},
\bibinfo{author}{\bibfnamefont{T.~C.}~\bibnamefont{Elston}},
\bibinfo{journal}{Biophys.~J.}
    \bibinfo{volume}{101} \bibinfo{year}{(2011)} \bibinfo{page}{144-150}.

\bibitem{Swie10}
\bibinfo{author}{\bibfnamefont{M.}~\bibnamefont{Swierczek}},
\bibinfo{author}{\bibfnamefont{E.}~\bibnamefont{Cieluch}},
\bibinfo{author}{\bibfnamefont{M.}~\bibnamefont{Sarewicz}},
\bibinfo{author}{\bibfnamefont{A.}~\bibnamefont{Borek}},
\bibinfo{author}{\bibfnamefont{C.~C.}~\bibnamefont{Moser}},
\bibinfo{author}{\bibfnamefont{P.~L.}~\bibnamefont{Dutton}},
\bibinfo{author}{\bibfnamefont{A.}~\bibnamefont{Osyczka}},
\bibinfo{journal}{Science}
    \bibinfo{volume}{329} \bibinfo{year}{(2010)} \bibinfo{page}{451-454}.

\bibitem{Sare15}
\bibinfo{author}{\bibfnamefont{M.}~\bibnamefont{Sarewicz}},
\bibinfo{author}{\bibfnamefont{A.}~\bibnamefont{Osyczka}},
\bibinfo{journal}{Physiol.~Rev.}
    \bibinfo{volume}{95} \bibinfo{year}{(2015)} \bibinfo{page}{219-243}.

\bibitem{Iver14}
\bibinfo{author}{\bibfnamefont{L.}~\bibnamefont{Iversen}},
\bibinfo{author}{\bibfnamefont{H.-L.}~\bibnamefont{Tu}},
\bibinfo{author}{\bibfnamefont{W.-C.}~\bibnamefont{Lin}},
\bibinfo{author}{\bibfnamefont{S.~M.}~\bibnamefont{Christensen}},
\bibinfo{author}{\bibfnamefont{S.~M.}~\bibnamefont{Abel}},
\bibinfo{author}{\bibfnamefont{J.}~\bibnamefont{Iwig}}
\bibinfo{author}{\bibfnamefont{et al.}},
\bibinfo{journal}{Science}
    \bibinfo{volume}{345} \bibinfo{year}{(2014)} \bibinfo{page}{50-54}.

\bibitem{Naka16}
\bibinfo{author}{\bibfnamefont{Y.}~\bibnamefont{Nakamura}},
\bibinfo{author}{\bibfnamefont{K.}~\bibnamefont{Hibino}},
\bibinfo{author}{\bibfnamefont{T.}~\bibnamefont{Yanagida}},
\bibinfo{author}{\bibfnamefont{Y.}~\bibnamefont{Sako}},
\bibinfo{journal}{Biophys.~Physbiol.}
    \bibinfo{volume}{13} \bibinfo{year}{(2016)} \bibinfo{page}{1-11}.

\bibitem{Crut12}
\bibinfo{author}{\bibfnamefont{J.~P.}~\bibnamefont{Crutchfield}},
\bibinfo{journal}{Nature Phys.}
    \bibinfo{volume}{8} \bibinfo{year}{(2012)} \bibinfo{page}{17-24}.

\bibitem{Penr70}
\bibinfo{author}{\bibfnamefont{O.}~\bibnamefont{Penrose}},
    \bibinfo{title}{Foundations of Statistical Mechanics. A Deductive Treatment.},
    \bibinfo{publisher}{Pergamon Press,~Oxford}, \bibinfo{year}{1970}.

\bibitem{Ande72}
\bibinfo{author}{\bibfnamefont{P.~W.}~\bibnamefont{Anderson}},
\bibinfo{journal}{Science}
    \bibinfo{volume}{177} \bibinfo{year}{(1972)} \bibinfo{page}{393-396}.

\bibitem{Zure03}
\bibinfo{author}{\bibfnamefont{W.~H.}~\bibnamefont{Zurek}},
\bibinfo{journal}{Revs.~Mod.~Phys.}
    \bibinfo{volume}{75} \bibinfo{year}{(2003)} \bibinfo{page}{715-765}.

\bibitem{Zure09}
\bibinfo{author}{\bibfnamefont{W.~H.}~\bibnamefont{Zurek}},
\bibinfo{journal}{Nature Phys.}
    \bibinfo{volume}{5} \bibinfo{year}{(2009)} \bibinfo{page}{181-188}.

\bibitem{Eldr72}
\bibinfo{author}{\bibfnamefont{N.}~\bibnamefont{Eldredge}},
\bibinfo{author}{\bibfnamefont{S.~J.}~\bibnamefont{Gould}},
    \bibinfo{title}{Punctuated equilibria: an alternative to phyletic gradualism.},
    \bibinfo{editors}{in T.~J.~M.~Schopf ed., Models in Paleobiology}, \bibinfo{page}{82-115},
    \bibinfo{publisher}{Freeman,~San Francisco} \bibinfo{year}{1972}.

\bibitem{Bak93}
\bibinfo{author}{\bibfnamefont{P.}~\bibnamefont{Bak}},
\bibinfo{author}{\bibfnamefont{P.}~\bibnamefont{Sneppen}},
\bibinfo{journal}{Phys.~Rev.~Lett.}
    \bibinfo{volume}{71} \bibinfo{year}{(1993)} \bibinfo{page}{4083-4086}.

\end{thebibliography}






\end{document}